\documentclass[twocolumn]{aastex631}

\begin{document}

\title{The One-hundred-deg$^2$ DECam Imaging in Narrowbands (ODIN): Survey Design and Science Goals}

\author[0000-0003-3004-9596]{Kyoung-Soo Lee}
\affiliation{Department of Physics and Astronomy, Purdue University, 525 Northwestern Ave., West Lafayette, IN 47906, USA}

\author[0000-0003-1530-8713]{Eric Gawiser}
\affiliation{Physics and Astronomy Department, Rutgers, The State University, Piscataway, NJ 08854}

\author[0000-0001-9521-6397]{Changbom Park}
\affiliation{Korea Institute for Advanced Study, 85 Hoegi-ro, Dongdaemun-gu, Seoul 02455, Republic of Korea}

\author[0000-0003-3078-2763]{Yujin Yang}
\affiliation{Korea Astronomy and Space Science Institute, 776 Daedeokdae-ro, Yuseong-gu, Daejeon 34055, Republic of Korea}

\author[0000-0001-5567-1301]{Francisco Valdes}
\affiliation{NSF's National Optical-Infrared Astronomy Research Laboratory, 950 N. Cherry Ave., Tucson, AZ 85719, USA}

\author[0000-0002-1172-0754]{Dustin Lang}
\affiliation{Perimeter Institute for Theoretical Physics, 31 Caroline Street North, Waterloo, ON N2L 2Y5, Canada}

\author[0000-0002-9176-7252]{Vandana Ramakrishnan}
\affiliation{Department of Physics and Astronomy, Purdue University, 525 Northwestern Ave., West Lafayette, IN 47906, USA}

\author[0009-0008-4022-3870]{Byeongha Moon}
\affiliation{Korea Astronomy and Space Science Institute, 776 Daedeokdae-ro, Yuseong-gu, Daejeon 34055, Republic of Korea}

\author[0000-0002-9811-2443]{Nicole Firestone}
\affiliation{Physics and Astronomy Department, Rutgers, The State University, Piscataway, NJ 08854}

\author[0000-0001-8227-9516]{Stephen Appleby}
\affiliation{Asia Pacific Center for Theoretical Physics, Pohang, 37673, Korea}
\affiliation{Department of Physics, POSTECH, Pohang 37673, Korea}

\author[0000-0003-0570-785X]{Maria Celeste Artale}
\affiliation{Departamento de Ciencias Fisicas, Universidad Andres Bello, Fernandez Concha 700, Las Condes, Santiago, Chile}

\author[0000-0002-1895-6639]{Moira Andrews}
\affiliation{Department of Physics and Astronomy, Purdue University, 525 Northwestern Ave., West Lafayette, IN 47906, USA}

\author[0000-0002-8686-8737]{Franz Bauer}
\affiliation{Instituto de Astrof{\'{\i}}sica, Facultad de F{\'{i}}sica, Pontificia Universidad Cat{\'{o}}lica de Chile, Campus San Joaquín, Av. Vicuna Mackenna 4860, Macul Santiago, Chile, 7820436}
\affiliation{Millennium Institute of Astrophysics, Nuncio Monse{\~{n}}or S{\'{o}}tero Sanz 100, Of 104, Providencia, Santiago, Chile}
\affiliation{Space Science Institute, 4750 Walnut Street, Suite 205, Boulder, Colorado 80301}

\author{Barbara Benda}
\affiliation{Physics and Astronomy Department, Rutgers, The State University, Piscataway, NJ 08854}

\author{Adam Broussard}
\affiliation{Physics and Astronomy Department, Rutgers, The State University, Piscataway, NJ 08854}

\author[0000-0001-6320-261X]{Yi-Kuan Chiang}
\affiliation{Institute of Astronomy and Astrophysics, Academia Sinica, Astronomy-Mathematics Building, Roosevelt Road, Taipei 10617, Taiwan} 

\author[0000-0002-1328-0211]{Robin Ciardullo}
\affiliation{Department of Astronomy \& Astrophysics, The Pennsylvania State University, University Park, PA 16802, USA}
\affiliation{Institute for Gravitation and the Cosmos, The Pennsylvania State University, University Park, PA 16802, USA}

\author[0000-0002-4928-4003]{Arjun Dey}
\affiliation{NSF's National Optical-Infrared Astronomy Research Laboratory, 950 N. Cherry Ave., Tucson, AZ 85719, USA}

\author{Rameen Farooq}
\affiliation{Physics and Astronomy Department, Rutgers, The State University, Piscataway, NJ 08854}

\author[0000-0001-6842-2371]{Caryl Gronwall}
\affiliation{Department of Astronomy \& Astrophysics, The Pennsylvania State University, University Park, PA 16802, USA}
\affiliation{Institute for Gravitation and the Cosmos, The Pennsylvania State University, University Park, PA 16802, USA}

\author[0000-0002-4902-0075]{Lucia Guaita}
\affiliation{Departamento de Ciencias Fisicas, Universidad Andres Bello, Fernandez Concha 700, Las Condes, Santiago, Chile}

\author[0000-0002-2073-5325]{Yun Huang}
\affiliation{Department of Physics and Astronomy, Purdue University, 525 Northwestern Ave., West Lafayette, IN 47906, USA}

\author[0000-0003-3428-7612]{Ho Seong Hwang}
\affiliation{Department of Physics and Astronomy, Seoul National University, 1 Gwanak-ro, Gwanak-gu, Seoul 08826, Republic of Korea}
\affiliation{SNU Astronomy Research Center, Seoul National University, 1 Gwanak-ro, Gwanak-gu, Seoul 08826, Republic of Korea}

\author[0009-0003-9748-4194]{Sang Hyeok Im}
\affiliation{Department of Physics and Astronomy, Seoul National University, 1 Gwanak-ro, Gwanak-gu, Seoul 08826, Republic of Korea}

\author[0000-0002-2770-808X]{Woong-Seob Jeong}
\affiliation{Korea Astronomy and Space Science Institute, 776 Daedeokdae-ro, Yuseong-gu, Daejeon 34055, Republic of Korea}

\author[0009-0002-6186-0293]{Shreya Karthikeyan}
\affiliation{Department of Astronomy, University of Maryland, College Park, MD 20742, USA}

\author[0000-0003-4770-688X]{Hwihyun Kim}
\affiliation{NSF's National Optical-Infrared Astronomy Research Laboratory, 950 N. Cherry Ave., Tucson, AZ 85719, USA}

\author[0009-0002-3931-6697]{Seongjae Kim}
\affiliation{Korea Astronomy and Space Science Institute, 776 Daedeokdae-ro, Yuseong-gu, Daejeon 34055, Republic of Korea}

\author[0000-0002-0905-342X]{Gautam R. Nagaraj}
\affiliation{Department of Astronomy \& Astrophysics, The Pennsylvania State University, University Park, PA 16802, USA}
\affiliation{Institute for Gravitation and the Cosmos, The Pennsylvania State University, University Park, PA 16802, USA}

\author[0000-0002-7356-0629]{Julie Nantais}
\affiliation{Departamento de Ciencias Fisicas, Universidad Andres Bello, Fernandez Concha 700, Las Condes, Santiago, Chile}

\author[0000-0001-9850-9419]{Nelson Padilla}
\affiliation{Instituto de Astronom\'ia Te\'orica y Experimental (IATE), CONICET-UNC, Laprida 854, X500BGR, C\'ordoba, Argentina}

\author[0000-0003-3095-6137]{Jaehong Park}
\affiliation{Korea Institute for Advanced Study, 85 Hoegi-ro, Dongdaemun-gu, Seoul 02455, Republic of Korea}

\author[0000-0001-8592-2706]{Alexandra Pope}
\affiliation{Department of Astronomy, University of Massachusetts, Amherst, MA 01003 }

\author[0000-0001-8245-7669]{Roxana Popescu}
\affiliation{Department of Astronomy, University of Massachusetts, Amherst, MA 01003 }

\author[0000-0002-5042-5088]{David Schlegel}
\affiliation{Physics Division, Lawrence Berkeley National Laboratory, 1 Cyclotron Road, Berkeley, CA 94720, USA }

\author[0009-0007-1810-5117]{Eunsuk Seo}
\affiliation{Department of Astronomy and Space Science, Chungnam National University, 99 Daehak-ro, Yuseong-gu, Daejeon, 34134, Republic of Korea}

\author{Akriti Singh}
\affiliation{Departamento de Ciencias Fisicas, Universidad Andres Bello, Fernandez Concha 700, Las Condes, Santiago, Chile}

\author[0000-0002-4362-4070]{Hyunmi Song}
\affiliation{Department of Astronomy and Space Science, Chungnam National University, 99 Daehak-ro, Yuseong-gu, Daejeon, 34134, Republic of Korea}

\author[0000-0001-6162-3023]{Paulina Troncoso}
\affiliation{Escuela de Ingenier\'{i}a, Universidad Central de Chile, Avenida Francisco de Aguirre 0405, 171-0614 La Serena, Coquimbo, Chile}

\author[0000-0003-4341-6172]{A.~Katherina~Vivas}
\affiliation{Cerro Tololo Inter-American Observatory/NSF's NOIRLab, Casilla 603, La Serena, Chile}

\author[0000-0001-6047-8469]{Ann Zabludoff}
\affiliation{Steward Observatory, University of Arizona, 933 North Cherry Avenue, Tucson AZ 85721}

\author[0000-0001-6455-9135]{Alfredo~Zenteno}
\affiliation{Cerro Tololo Inter-American Observatory/NSF's NOIRLab, Casilla 603, La Serena, Chile}

%\author{the ODIN Collaboration}
%\affiliation{}

%\author{everyone in the odin collaboration}
%\collaboration{20}{(AAS Journals Data Editors)}



\begin{abstract}

%250 word limit for ApJ - currently 244 words 

We describe the survey design and science goals for ODIN (One-hundred-deg$^2$ DECam Imaging in Narrowbands), a NOIRLab survey using the Dark Energy Camera (DECam) to obtain deep (AB$\sim$25.7) narrow-band images over an unprecedented area of sky.  The three custom-built narrow-band filters, $N419$, $N501$, and $N673$, have central wavelengths of 419, 501, and 673 nm and respective full-width-at-half-maxima of 7.2, 7.4, and 9.8 nm, corresponding to Ly$\alpha$ at $z=$2.4, 3.1, and 4.5 and cosmic times of 2.8, 2.1, and 1.4 Gyr, respectively.  When combined with even deeper, public broad-band data from Hyper Suprime-Cam, DECam, and in the future, LSST, the ODIN narrow-band images will enable the selection of over 100,000 Ly$\alpha$-emitting (LAE) galaxies at these epochs.    ODIN-selected LAEs will identify protoclusters as galaxy overdensities, and the deep narrow-band images enable detection of highly extended Ly$\alpha$ blobs (LABs).  Primary science goals include measuring the clustering strength and dark matter halo connection of LAEs, LABs, and protoclusters, and their respective relationship to filaments in the cosmic web.  The three epochs allow the redshift evolution of these properties to be determined during the period known as Cosmic Noon, where star formation was at its peak.
The narrow-band filter wavelengths are designed to enable interloper rejection and further scientific studies by revealing  [O~{\sc ii}] and [O~{\sc iii}] at $z=0.34$,  Ly$\alpha$ and He~{\sc ii}~1640 at $z=3.1$, and Lyman continuum plus Ly$\alpha$ at $z=4.5$. Ancillary science includes similar studies of the lower-redshift emission-line galaxy samples and investigations of nearby star-forming galaxies resolved into numerous [O~{\sc iii}] and [S~{\sc ii}] emitting regions.  

\end{abstract}

\section{Introduction}
In the hierarchical theory of structure formation,   initial density fluctuations grow via gravitational instabilities and form filaments, sheets, voids, and collapsed halos \citep{bond96}. These features are collectively referred to as large-scale structures (LSS). In this picture, a galaxy's fate is primarily determined by the surrounding LSS, which controls the rate and the timeline of gas accretion and feedback processes, as well as the likelihood of galaxy-galaxy interaction. 

Observational evidence supporting this expectation is abundant. 
At physical scales of $\approx$100's~kpc, galaxies with a close companion tend to have enhanced star formation rates (SFRs) and follow a different mass-metallicity relation than that for isolated `field' galaxies \citep[e.g.,][]{ellison08,hwang11}. 
At cluster scales ($\lesssim 1$~Mpc), the shape of the galaxy stellar mass function in cluster environment differs from that of average field \cite[e.g.,][]{baldry06} in that fewer low-mass star-forming galaxies and a higher fraction of quenched galaxies are present. Similar trends persist in group-scale environments and in high-redshift ($z\approx1$) clusters \citep{yang09, peng10,vanderburg20}. Such environmental dependence may begin at $z\gtrsim2$ \citep{lemaux22}, i.e., well before a large cosmic structure becomes fully virialized. 
At the largest scales (up to $\approx$10~Mpc comoving), the degree of galaxy clustering is a strong function of stellar mass, SFR, luminosity, and spectral type \citep[e.g.,][]{norberg02,adelberger05,giavalisco01,lee06}, suggesting that galaxies with larger stellar masses, higher SFR, and/or redder colors are hosted by more massive halos than their less massive, bluer counterparts. 

At $z\gtrsim 2$, when the global star formation activity reached its peak \citep{madau14}, the role of the LSS environment in galaxy formation remains under-explored.  Many of the strong spectral features redshift to the infrared wavelength, making ground-based investigations difficult.
%beyond the observational reach from ground-based facilities.
Moreover, severe cosmological dimming requires long integrations on large-aperture telescopes to detect continuum emission and measure redshift even for relatively bright galaxies, rendering a  complete spectroscopic survey unfeasible. While the {\it James Webb Space Telescope} (JWST) and 30m-class telescopes will certainly alleviate these challenges, the small field of view probed by these facilities (typically, a few arcminutes on a side)  is better suited for detailed follow-up studies of a few interesting systems rather than for general exploratory surveys large enough to probe the LSS. 

At high redshift, Ly$\alpha$ emission provides one of the primary windows into the high-redshift universe \citep[see review by][and references therein]{ouchi20}, as it traces ionized and/or excited gas from star formation, black hole activity, and the gravitational collapse of dark matter halos. 
The Ly$\alpha$ emission strength from $z>2$ galaxies is inversely correlated with stellar mass and internal extinction, meaning that Ly$\alpha$ emitters (LAEs) tend to have younger ages and lower SFRs than continuum-selected galaxies \citep[e.g., ][]{gawiser06,guaita11}.
%Existing studies show that a great majority of Ly$\alpha$-emitting galaxies (LAEs) tend to have lower stellar masses, lower SFRs, and younger ages \citep{gawiser06,guaita11}. 
They are also less dusty than any other known galaxy population \citep{weiss21}. 
Given the relative ease of narrow-band imaging compared to spectroscopy and the reduced projection effects of narrow- versus broad-band selection, LAEs are the most efficient tracer of the underlying matter distribution at high redshift. 
The utility of LAEs 
%in tracing large-scale structure  in the distant universe 
for this purpose
has been known for two decades \citep[e.g.,][]{hum96,ouchietal03,kovacetal07,gawiser07}. 
Angular clustering measurements suggest that LAEs are hosted by moderate-mass halos \citep{gawiser07,guaita10,lee14}. These traits give LAEs the lowest clustering bias among visible tracers of the underlying matter distribution.
%, providing a unique view into the LSS of the universe to study galaxy formation and cosmology.  
Finally, Ly$\alpha$ mapping via narrow-band imaging offers an efficient alternative for collecting large galaxy samples in a well-defined volume with minimal contamination from foreground sources. The redshift precision, $\Delta z\sim 0.04-0.06$ is several times better than that afforded by 
the best photometric redshifts that can be derived using broad-band filters 
%the state-of-the-art photometric redshift 
($\Delta z\sim 0.1-0.2$ at $z>2$). 

%- a paragraph about known protoclusters as LAE overdensities.

%- motivation of the ODIN survey

%- make a connection to the LSST/Rubin era

%- summary of rest of this paper 
This paper is organized as follows. In Section~\ref{sec:odin_science}, we describe the key science goals of the ODIN survey and provide a broad context and an overview of each topic. Section~\ref{sec:survey_design} summarizes the survey parameters, detailing the filter transmission, and the key characteristics of the survey fields; additionally, we justify the imaging depths and present the expected outcomes upon completion of the survey. In Section~\ref{sec:obs}, we discuss the general procedures guiding the data acquisition and reduction, and in Section~\ref{sec:external_data}, we describe the survey's existing and future data. Finally, in Section~\ref{sec:other_applications}, we stress the legacy value of the ODIN data and the ODIN filters in advancing Galactic science. We adopt a $\Lambda$CDM concordance cosmology with $h=0.7$, $\Omega_m=0.27$, and $\Omega_\Lambda=0.73$ and use comoving distance scales unless noted otherwise. All magnitudes are in the AB scale \citep{oke83}. 

\section{Primary ODIN Science Goals}\label{sec:odin_science}

The One-hundred-deg$^2$  DECam Imaging in Narrowbands (ODIN) is designed to enable a multitude of science goals but is primarily geared towards understanding the formation and evolution of galaxies in the distant universe. 
ODIN  probes the large-scale structure in  three narrow cosmic slices straddling the epoch of ``Cosmic Noon'' using custom-built narrow-band filters.  These filters 
sample redshifted Ly$\alpha$ at cosmic ages of 2.8, 2.1, and 1.4~Gyr ($z=2.4$, 3.1, and 4.5), strategically placed to study the redshift evolution of galaxies and cosmic structures. 
When combined with the upcoming deep wide-field surveys to be conducted with the Rubin Observatory's Legacy Survey of Space and Time (LSST), the Nancy Grace Roman telescope, and Euclid, ODIN has the potential to reveal the full diversity of astrophysical phenomena in the context of their LSS, as well as to measure the topology of LSS thereby constraining cosmological parameters. 
In what follows, we describe the primary goals of the ODIN survey.

\subsection{Protoclusters, cosmic web, and galaxy inhabitants}

Protoclusters are progenitors of the most massive gravitationally bound structures, i.e., clusters of galaxies. As such, protoclusters allow us to directly witness the formation and evolution of galaxies, thereby shedding light onto the physical processes that led to their early accelerated formation followed by strong and swift quenching that occurred at high redshift \citep[e.g.,][]{stanford98,blakeslee03,thomas05,eisenhardt08,mancone10}.

%While it is widely accepted that cluster galaxies underwent an early accelerated formation followed by strong and swift quenching that occurred at high redshift \citep[e.g.,][]{stanford98,blakeslee03,thomas05,eisenhardt08,mancone10}, the details of how and when the local environment influenced their formation and evolution can only be obtained by tracing the galaxy growth in young protoclusters. 

The limitations of studying protoclusters have been muti-fold: first, clean and statistically robust samples of protoclusters have been difficult to obtain. The lack of readily identifiable signatures -- such as a hot intracluster medium and/or a concentration of quiescent galaxies --  in young, yet-to-be-virialized structures of mostly star-forming galaxies leads to a strong reliance on spectroscopy for finding protoclusters. %The faint nature of high-redshift sources requires deep spectroscopy which often introduces a  confirmation bias for line-emitting members. 
Photometric redshifts from broad- and intermediate-band filters can help with selection efficiency \citep[e.g.,][]{chiang14}, but few extragalactic fields have the data necessary to produce photo-z's with the requisite $\Delta z\approx 0.1$ precision \citep[e.g.,][]{shi19,huang22}.

Second, protoclusters subtend tens of arcminutes in the sky \citep{muldrew15,chiang13} making member identifications challenging. Even if the general region of a protocluster were targeted, blind spectroscopy would be dominated by fore- and background interlopers. 
 Finally, massive clusters are very rare. The most massive, Coma-like clusters ($M_{\rm today} \gtrsim 10^{15}M_\odot$) have a space density of $\approx$180~Gpc$^{-3}$ \citep{reiprich02}, corresponding to $\approx$2.3~deg$^{-2}$ per unit redshift. 
Most surveys are ill-suited for discovering such sparsely distributed structures.

These difficulties explain why there are very few well-characterized protocluster systems \citep[e.g.][]{dey16,cucciati18}. Despite the rapidly growing number of protoclusters and their candidates \citep[e.g.][]{planck_overdensity,toshikawa18}, there is a critical need for reliable markers or `signposts' of protoclusters.  
Though possible candidates are numerous \citep[see][and the references therein, but the candidates include radio galaxies, quasi-stellar objects, and Ly$\alpha$ nebulae]{overzier16}, the extent of their success varies even when it can be measured. Moreover, the intersection of these markers, their respective selection biases, and their duty cycle (i.e., the typical timescale in which they stay visible) are currently unknown. Uniformly selected large samples of protoclusters are needed to address these questions. 

ODIN uses LAE overdensities as tracers of massive cosmic structures. This is motivated by the fact that LAE samples 
%have the smallest clustering bias of any commonly used tracer of high-z structures
%%are the least biased galaxy tracers 
%(see Section~\ref{sec:intro}), 
suffer from much weaker projection effects than continuum-selected galaxies due to the narrow redshift interval and offer high sky density with contiguous coverage that cannot easily be achieved with spectroscopy. Additionally,  studies of the few well-characterized protoclusters, such as SSA22, {\it Hyperion}, and PC217.9+32.3 \citep{steidel99,lee14,dey16,topping18,huang22}, suggest that LAEs trace the overall galaxy overdensity structures quite well, highlighting the systems' filament-like structures and knots. 

In this context, one of the immediate goals of  ODIN  in advancing protocluster science are to procure a statistically complete sample of protoclusters via the identification of low-mass, low-luminosity galaxies, and to explore the intersectionality with structures of dusty star-forming galaxies \citep[e.g.,][]{casey16,planck_overdensity}, Lyman-break galaxies, extended Ly$\alpha$ nebulae, and radio galaxies. The survey volume and the target fields (Section~\ref{sec:survey_design}) are chosen to allow robust, statistically significant comparisons of the samples. A first analysis of how the ODIN-selected Ly$\alpha$ nebulae are located in the context of LAE-traced cosmic filaments and protoclusters is provided in \citet{ramakrishnan23}. The ODIN fields coincide with the regions with the deepest {\it Spitzer} IRAC data (Section~\ref{subsec:spitzer}), which 
will enable a systematic search for the progenitors of giant cluster ellipticals, some of which are expected to have already been formed at high redshift  \citep[e.g.,][]{thomas05,mei06}. Deeper IR data from Roman in the next decade will push the limit further.

\subsection{Extended Ly$\alpha$ nebulae}

%- describe what they are and their characteristics

Extended Ly$\alpha$ nebulae, also known as Ly$\alpha$ ``blobs'' (LABs), emit Ly$\alpha$ radiation with total line luminosities of $10^{43}-10^{44}$~erg~s$^{-1}$ on physical scales of 20--100~kpc \citep[e.g.,][]{francis96,steidel00,matsuda04,saito06,yang09,matsuda11}. This far exceeds the size of a single galaxy. 

%- possible powering mechanisms
Given their extreme physical properties, the physical nature of LABs has astrophysical and cosmological implications. Some blobs are clearly produced by photoionization from intense radiation produced by embedded active galactic nuclei (AGN), which themselves may or may not be visible at optical wavelengths \citep[e.g.,][]{kollmeier10,overzier13,yang14a}. The recombination of ionized gas results in  Ly$\alpha$ photons, which resonantly scatter and illuminate the surrounding H~{\sc i} gas. In a few spectacular cases around luminous QSOs \citep[e.g.,][]{borisova16,arrigoni19} or multiple AGN, the scattered Ly$\alpha$ emission reveals cosmic filaments connected to the source \citep[e.g.,][]{cantalupo14,umehata19}. Alternatively, Ly$\alpha$ photons produced in the star-forming regions of embedded galaxies may scatter outward and power LABs \citep[e.g., ][]{laursen07,cen13,chang23}. Shock heating of the gas in the circumgalactic medium by starburst-driven superwinds has also been proposed as a possible  mechanism \citep[e.g., ][]{taniguchi00}.

Much attention has been given to LABs as potential signposts of gas accretion along cosmic filaments. As H~{\sc i} gas falls into the potential well of a dark matter halo and is heated to $\gtrsim 10^4$~K, a large fraction of the gravitational cooling radiation is expected to be emitted in Ly$\alpha$  \citep{haiman00,fardal01}.  Although the tight correlation between stellar mass and SFRs, often referred to as the star-forming galaxy `main sequence' \citep[e.g.,][]{noeske07}, implies that the dominant mode of galaxy growth is through a relatively smooth and prolonged inflow of gas, direct and clear evidence for this inflow has been challenging to obtain. 

%- recent findings that they may trace overdense structures, or cold accretion

Recent ultradeep integral field observations suggest that some LABs may indeed be powered by gravitational cooling. \citet{daddi21} showed that the morphology and kinematics of the extended Ly$\alpha$ emission around a galaxy group at $z=2.91$ are consistent with gas inflow along multiple cosmic filaments converging onto the group. By studying extended Ly$\alpha$ emission around a handful of protoclusters or groups of galaxies, \citet{daddi22} reported tentative evidence that  LAB luminosity correlates with dark matter halo mass (and thus with expected baryon accretion rate). These studies highlight the fact that direct detection of cold gas accretion near massive halos may be at the edge of the current observational reach. 
%With the 30m-class telescopes on the horizon, the next step forward will be to secure a sufficiently large number of LABs with well-characterized environmental information. 

Indeed, many known blobs are associated with galaxy overdensities \citep[e.g., ][]{shi19} or lie near what appears to be a filament traced by galaxies \citep[e.g., ][]{erb11}. \citet{prescott08} surveyed the region surrounding a very luminous, mid-IR-detected blob \citep{dey05} and reported an overdensity of galaxies \citep[also see][]{xue17}. A Ly$\alpha$ survey around a known protocluster at $z=3.09$, SSA22, revealed multiple Ly$\alpha$ blobs \citep{steidel00} residing in regions that appear to be converging filaments \citep{matsuda04}. \citet{badescu17} noted that blobs tend to reside in the vicinity of protoclusters identified as LAE overdensities \citep[also see][]{ramakrishnan23}. When combined with their high cosmic variance \citep{yang10}, the authors speculate that blobs may represent infalling proto-galaxy-groups.

%- current limitations in making progress and what ODIN will do

To place definitive constraints on the physical origin(s) of LABs and their connection to cosmic structures, large and statistically complete samples of LABs are crucial. By detecting over 1,000 bright LABs at three cosmic epochs, ODIN will produce the largest blob sample to date and enable robust measurements of the sizes, and luminosities, and redshift evolution of these objects. Auto- and cross-correlation function measures will inform us of typical masses of LAB host halos and their interrelation with known cosmic structures. When combined with the ODIN protocluster samples, it will be possible to determine whether or not LABs are reliable tracers of protoclusters, which in turn constrains the duty cycle of the LAB phenomenon.

\subsection{Physical properties of LAEs via LFs, SEDs,  and Clustering}
LAEs represent a low-dust, high specific star formation rate stage of galaxy evolution, possibly consistent with galaxies undergoing initial starbursts \citep{partridge67, gawiser07}, 
making it important to fully characterize their properties.
By detecting tens of thousands of LAEs situated in the wide and deep imaging fields in the equatorial and Southern sky, ODIN aims to greatly increase our knowledge of these galaxies' physical properties.  

With the rapid increase in sophistication of cosmological hydrodynamic simulations and semi-analytic models of galaxy formation, it has become possible to test theoretical predictions for the distribution of directly observed galaxy properties including their continuum and emission-line luminosities and colors.  LAEs offer a unique probe of the low-mass bulk of the high-redshift galaxy distribution, but connecting them with these models requires matching the observed distribution of Ly$\alpha$ emission strengths i.e., the Ly$\alpha$ luminosity function (LF), which report the number density of LAEs as a function of their Ly$\alpha$ luminosity \citep[e.g.,][]{gronwalletal07, ciardulloetal12,sobral18}; the large statistical samples enabled by ODIN's combination of depth and area and the strong interloper rejection enabled by deep photometry in multiple broad-band filters will yield improved LF measurements at each redshift.  

%Existing studies show that LAEs tend to have low stellar masses, young ages, and high ratios of star formation rate to stellar mass \citep{gawiser06,gronwalletal07,guaita11}; they are also less dusty than any other known galaxy population  \citep{gawiser07,guaita10,lee14,kusakabe18,weiss21}. 
Deep rest-UV-through-near-IR broadband photometry of LAEs available in the ODIN fields (Table~\ref{tab:fields}) will enable improved Spectral Energy Distribution (SED) fitting to measure SFR, stellar masses, and dust extinction.  \citet{vargasetal14} found that, with broad-band photometry significantly deeper than the narrow-band photometry used to select LAEs, it was not necessary to stack LAE SEDs in order to obtain robust parameter fits.  
Sufficiently deep 
$u$-band-through-IRAC 
photometry is available in the CANDELS \citep{Nayyeri2017:CANDELS_COSMOS, Merlin2021:ASTRODEEP_GOODSS} and COSMOS2020 \citep{Weaver2021:COSMOS2020} catalogs.  
Shallower $ugrizyK$+IRAC SEDs are available for stacked SED analysis of $>$30,000 LAEs in the 24 deg$^2$ field of SHELA \citep{Papovich2016:SHELA,wold19,Stevans2021:NEWFIRM}.    

\begin{figure*}[ht!]
\centering
\includegraphics[scale=0.05]{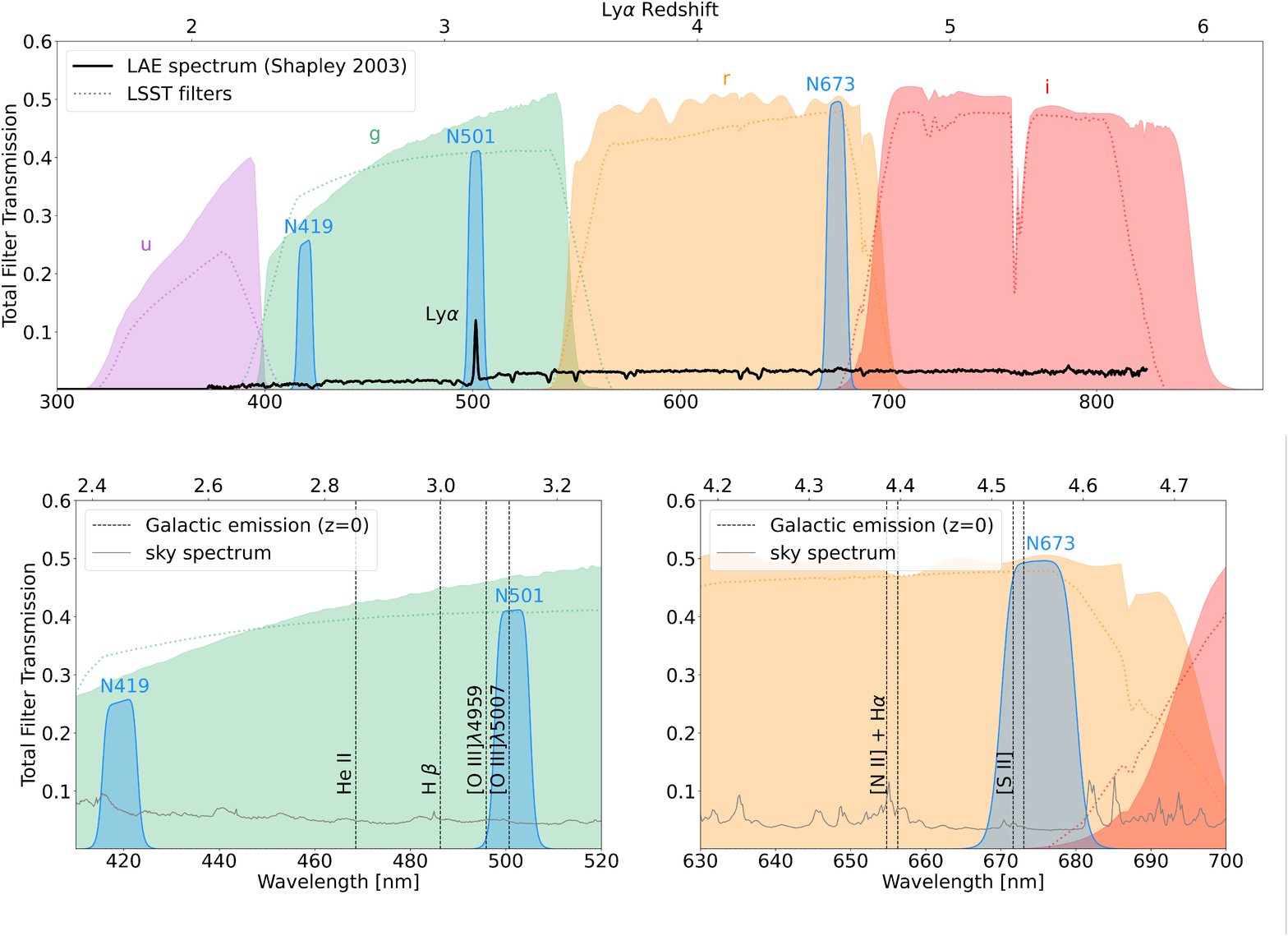}
\caption{
{\it Top:} The ODIN narrowband filters made for DECam (blue shaded regions; from left to right, $N419$, $N501$, and $N673$, respectively) together with the transmission curves for the broad-band filters from Subaru HSC ($gri$) and  CFHT/MegaCam ($u$). Future LSST passbands for the same broad-bands are shown as dotted lines. These imaging data are combined to isolate redshifted Ly$\alpha$ emission falling into the narrowband filters as illustrated by the template LAE spectrum (black) at $z=3.1$. The Ly$\alpha$ redshift range is shown at the top. {\it Bottom:} Zoom-in views of the three ODIN filters. Overlaid are strong Galactic emission lines (vertical dotted lines) and the sky spectrum (grey curves). $N501$ and $N673$ sample the [O\,{\sc iii}]$\lambda$5007 and [S\,{\sc ii}] emission from Galactic sources and nearby galaxies.     
}
\label{fig:filters}
\end{figure*}

Clustering analysis 
allows us to connect LAEs to their host dark matter halos to answer key questions about galaxy evolution. Extended Press-Schechter theory \citep{sheth01} and cosmological $N$-body simulations predict the clustering bias of dark matter halos as a function of halo mass and redshift, and this has been used to learn that LAEs are hosted by dark matter halos with masses  $\sim10^{11}$M$_\odot$ \citep{gawiser07, guaita10, kusakabe18}.
But the largest sample of LAEs analyzed so far consists of 
4,400 LAEs over 19 deg$^2$ at $z=2.2$ \citep{Ono:2021} 
%1250 LAEs over 1 deg$^2$ at $z=2.2$ \citep{kusakabe18} and equivalent numbers over 20 deg$^2$ at $z=5.7$ and $z=6.6$ 
combined from SILVERRUSH \citep{ouchi18} and CHORUS \citep{Inoue2020:CHORUS}.  
Clustering analyses over areas
of 0.25--1.6 deg$^2$ have also been performed at $z=2.1$ \citep[250 LAEs,][]{guaita10}, $z=3.1$ (356 LAEs, \citealt{Ouchi:2010} and 162 LAEs, \citealt{gawiser07}),
$z=3.3$ \citep[959 LAEs,][]{Ono:2021},
$z=4.5$ \citep[151 LAEs,][]{kovacetal07}, and 
$z=4.9$ \citep[349 LAEs,][]{Ono:2021},
$z=5.7$ \citep[2881 LAEs,][]{ouchi18}, and
$z=6.6$ \citep[680 LAEs,][]{ouchi18}. 

The combination of larger area reducing cosmic variance \citep[see Figure~5 in][]{kusakabe18} and increased statistics reducing Poisson-like errors on correlation function measurement will allow us to improve the constraints on the bias of dark matter halos hosting LAEs.   
 Following \citet{ouchi18}, we can include halo occupation distribution parameters when modeling the observed angular correlation function to determine the fractional occupation of dark matter halos by LAEs as a function of halo mass and the number of LAEs that are satellite galaxies in more massive halos.

\subsection{Reconstructing the expansion history of the universe}
%\textcolor{blue}{This section is a work in progress courtesy of Stephen Appleby and Changbom Park.}
%The cosmic web revealed by the LAE distribution in our narrow-band images will provide a novel probe of the expansion history  at $z>2$,  where dark energy is not expected to affect the expansion but modified gravity or other new physics (e.g., early dark energy) could be revealed. Our survey will produce maps of LAEs at $z\sim 2.4$, 3.1, and 4.5 over large contiguous areas (up to $r$=0.5~Gpc) allowing us to trace a wealth of information on the LSS. The topology of LSS -- i.e., connectivity of dense and underdense structures measured via a genus statistic -- is conserved on large scales including quasi-linear scales ($r\gtrsim 10$~Mpc comoving) to high accuracy ($\lesssim$2\%) and can be used to constrain cosmological parameters (Park \& Kim 2010; Appleby+2017, 2020). These features are relatively insensitive to LAE bias, which may change with redshift due to selection effects and galaxy evolution.  

 The large-scale distribution of LAEs revealed by ODIN will provide a unique opportunity to measure the expansion history at $z > 2$, and hence test the standard model of cosmology. Recently found tensions between cosmological parameters inferred from early and late Universe data hint at new physics beyond the $\Lambda$CDM model, which could include departures from General Relativity or the presence of additional energy density components such as early dark energy. LSS surveys can be used to reconstruct the expansion history of the Universe by utilizing standard rulers or populations. 

One example is the amplitude of the genus of the matter density. The genus is a member of the Minkowski Functionals and is a topological quantity that provides a measure of the connectivity of the matter distribution. The amplitude of the genus is relatively insensitive to redshift-dependent tracer bias and non-Gaussianities induced by gravitational collapse on quasi-linear $\sim {\cal O}\left(10 \, {\rm Mpc}\right)$ comoving scales. Hence, this statistic provides a relatively pure measurement of the connectivity of the initial state of the dark matter field and should be conserved with redshift to high accuracy $\sim {\cal O} (1\%)$. 
%However, 
Spurious evolution in the genus amplitude would be observed if an incorrect cosmology is selected.
%to infer the comoving distance between the observer and LSS data. 
Consequently, by measuring this quantity at different redshifts, one can determine the distance-redshift relation that minimizes the evolution \citep{Park:2009ja}.

Another possibility is to use the shape of the two-point correlation function and apply the extended Alcock-Paczynski test \citep{Park:2019mvn,Dong:2023jtk} to the ODIN  data. As the shape of the two-point correlation function of cosmic objects is insensitive to the growth of structure and their bias, it is a good `standard shape' and can be used to measure the expansion history of the universe. Recently, the method has been applied to the low-redshift SDSS survey data, and a tight constraint on the dark energy equation of state parameter $w =-0.903\pm 0.023$ was obtained for a flat $w$CDM universe, which is $4.2\sigma$ away from the $\Lambda$CDM model \citep{Dong:2023jtk}. The analysis can be applied to the higher redshift universe revealed by ODIN, to explore the possibility of potential non-standard dark energy signatures, and more generally test if the observed universe is consistent with a decelerating epoch as predicted by the current standard cosmology.

ODIN   will produce maps of LAEs at $z = 2.4$, 3.1, and $4.5$ over large contiguous areas, which is ideal for studying LSS using these methods. Although we expect the LAE bias to be redshift-dependent due to selection effects and galaxy evolution, the genus amplitude should be practically insensitive to these artifacts. The genus of two-dimensional slices of the SDSS-III Baryon Oscillation Spectroscopic Survey \citep[][]{2000AJ....120.1579Y} galaxy distribution at $z \sim 0.5$ has previously been measured \citep{Appleby:2017ahh,Appleby:2021lfq}, and the genus extracted from the LAEs can be directly compared to these low-redshift results. 

To perform this analysis, we must first understand how LAEs are biased relative to the underlying  matter distribution. This can be determined using cosmological hydrodynamic simulations. Horizon Run 5 \citep[HR5:][]{Lee_2021,Park_2022} is a cosmological hydrodynamical simulation with a simulation box size of $\sim 1 \, {\rm cGpc}^{3}$ and a high-resolution cuboid zoom-in region of $1049 \times 119 \times 127 \, ({\rm cMpc})^{3}$, where the spatial resolution reaches down to $\sim 1 \, {\rm kpc}$. The geometry of the zoom-in region is designed to be optimized for generating mock lightcone data for deep field surveys such as ODIN. Mock LAE data can be constructed using semi-analytic prescriptions, and the relation between LAEs and dark matter fields can be directly compared on quasi-linear scales using snapshot data from the simulation.

\section{The ODIN Survey Design} \label{sec:survey_design}

\subsection{ODIN Narrow-band Filters}

In Figure~\ref{fig:filters}, we show the three ODIN narrow-band filters made for the Dark Energy Camera. The same figure also shows the transmission curves for the broad-band filters we use in conjunction to make LAE selections. These are the Subaru Hyper Suprime-Cam filters ($grizy$), the CFHT MegaCam $u$-band filter, and (in the near future) the LSST $ugrizy$ filters. All data represent the total throughput including filter transmission, optics, detector, and mirror. The basic filter characteristics are summarized in Table~1 together with the expected redshift selection function of each filter, peaking at $z\approx2.4$, 3.1, and 4.5, for the $N419$, $N501$, and $N673$ filters, respectively.

\begin{figure*}
\centering
\includegraphics[scale=0.7]{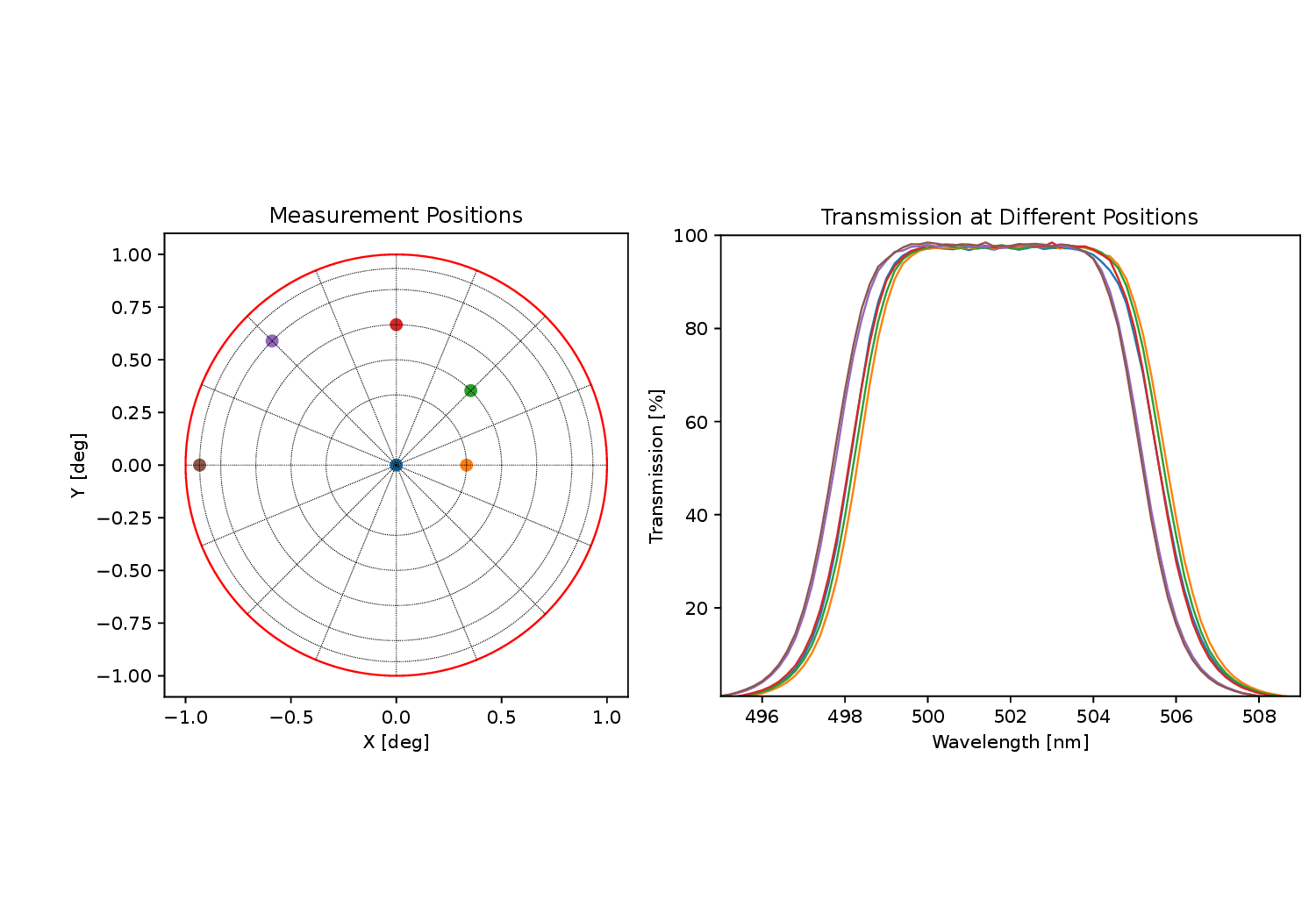}
\vspace{-0.8in}
\caption{
{\it Left panel:} The focal plane of DECam is shown as a red circle. Of the 49 positions at which the measurements were made, we only show the $N501$ filter transmission at the six positions indicated by the colored circles. Variations in the transmission curve are primarily a function of distance from the center. {\it Right panel:} While the transmission curve shifts blueward by a few angstroms near the edge of the field, it remains relatively constant over the bulk of the DECam focal plane. The behavior of $N419$ and $N673$ filters is similar to that shown in this figure.
}
\label{fig:transmission}
\end{figure*}
The ODIN filters are designed for dual use, i.e., to promote both high-redshift and Galactic science. 
%The $N501$ filter can effectively isolate  [O~{\sc iii}]$\lambda5007$  at $\Delta v=\pm 
%500$~km~s$^{-1}$  from [O~{\sc iii}]$\lambda4959$. 
At zero velocity, the filter transmission at 4959~\AA\ and 5007~\AA\ is 10\% and 99\%, respectively. At $\Delta v=\pm 500$~km~s$^{-1}$, the transmission changes to 5\% (25\%) and 97\% (99\%). Similarly, The $N673$ filter samples [S~{\sc ii}] emission at $|v|\leq 500$~km~s$^{-1}$ at a nearly constant transmission. 

A combination of $N673$ and $N501$ can identify  [O~{\sc iii}] emitters at $z=0.34$ masquerading as $N673$-detected LAEs (N. Firestone et al. in preparation). For $N501$-detected LAEs at $z=3.1$, $N673$ samples He~{\sc ii}1640 emission to constrain their average stellar population properties via stacking analysis. For $N673$-detected LAEs at $z=4.5$, Lyman continuum radiation that leaks out can be detected in the $N501$ filter.  
%[cite Lucia's thesis for the last two ideas.]

All three filters are circular in shape with a diameter of 620~mm and are designed to have a transmission curve as close as possible to a top hat in shape. The post-production lab measurements were made in a parallel beam with 5$^\circ$ angle of incidence at the Asahi Spectra Ltd. facility on 49 points distributed along five concentric circles across the focal plane to characterize the variation of transmission. The expected transmission at normal incidence (f/3.6) is then computed based on these measurements. In Figure~\ref{fig:transmission}, we show the results at five different angular distances from the field center. As one moves farther away from the center of DECam's 1.1$^\circ$ circular field,  the central wavelength tends to shift blueward, and the transmission increasingly deviates from being top-hat in shape.

%On-site transmission measurements at CTIO \textcolor{blue}{be more specific about the procedure} appear consistent with the lab results; the transmission variations across the focal plane are yet to be validated and quantified via observations of bright stars. 

The change in filter transmission as a function of field position is expected to have a negligible impact on our science goals. As we discuss in Section~\ref{subsec:dither}, the dither patterns for most ODIN fields are designed to minimize transmission variations by observing the same objects at many locations within the focal plane.  Doing so ensures that, throughout a given field, the effective transmission is as close as possible to a weighted average of the curves shown in Figure~\ref{fig:transmission}. Upon observational validation of these fluctuations, it may be possible to utilize the transmission variation to perform a single-filter `tomography' and determine a more precise redshift of bright LAEs \citep{Zabl:2016}. 

\begin{table}
\caption{Filter Summary}
\vspace{-0.25in}
\begin{center}
\begin{tabular}{lcccccc}
\hline\hline
Filter & $\lambda_C$    & $\lambda_{l}^\ast$  & $\lambda_{u}^\ast$  &  FWHM & $z_C$ & $\Delta z$\\
&  [nm] & [nm] & [nm] & [nm] & & \\ 
\hline
$N419$ & 419.3 & 415.5 & 423.0 & 7.4 & 2.449 & 0.061 \\
$N501$ & 501.4 & 497.6 & 505.2 & 7.2 & 3.124 & 0.059 \\
$N673$ & 675.0 & 670.0 & 680.0 & 9.8 & 4.553 & 0.081 \\
\hline
\hline
\end{tabular}
\end{center}
{\small
$^\ast$$\lambda_l$ and $\lambda_u$ give the endpoints of each filter's full width at half maximum (FWHM). $z_C$ and $\Delta z$ represent the central wavelength and the FWHM of each filter in redshift space, respectively.   \\
%$^\dagger$
}
\end{table}
%\begin{itemize}
%    \item show filter transmission variations across the DECam focal point?
%\end{itemize}

\subsection{ODIN Survey Fields} 

\begin{table*}
\caption{Summary of ODIN Fields and  Available Data}
\begin{center}
\begin{tabular}{llccll}
\hline
Field &  Field Center & Area &  Pointing Pattern & Notable Facts \\
 & (J2000)  &  [deg$^2$] &  &  \\
\hline
ELAIS-S1 & 9.45$^\circ$, $-44.00^\circ$ & 10 &  Two Rings & DDF\\ 
SHELA$^\ast$ & $20.5^\circ$, $0.00^\circ$ &  24 &   Spiral & HETDEX, DESI access\\
XMM-LSS & $35.71^\circ$, $-4.75^\circ$ & 10 &  Two Rings &  DDF, SSP, DESI access \\
CDF-S & $53.13^\circ$, $-28.10^\circ$ & 10  & Two Rings & DDF, EDF\\
EDF-Sa & $58.90^\circ$, $-49.32^\circ$ & 10  &  Two Rings & EDF, SPHEREx DF\\
EDF-Sb & $63.60^\circ$, $-47.60^\circ$ & 10 &   Two Rings & EDF, SPHEREx DF\\
E-COSMOS & $150.10^\circ$, $2.18^\circ$ & 10 &  Two Rings & DDF, SSP, DESI access\\
Deep2-3 & $352.1^\circ$, $-0.28^\circ$ & 7 &  Spiral & SSP, DESI access\\
\hline
{\bf Total} & - & {\bf 91} &  \\
\hline\hline
Survey & BB Filters & &  $5\sigma$ Point Source Sensitivity  \\
\hline
SSP Deep$^\dagger$ & $ugrizy$ & - &  27.1, 27.5, 27.1, 26.8, 26.3, 25.3\\
%SHELA$^\ast$ & $ugrizy$ & - & \\
LSST DDF Y1 & $ugrizy$ & - &   26.7, 27.9, 28.1, 27.4, 26.6, 25.3\\
%LSST DDF Y3 & $ugrizy$ & -   & 27.3, 28.5, 28.7, 28.0, 27.2, 25.9\\
LSST DDF Y10 & $ugrizy$ & -   & 28.0, 29.2, 29.4, 28.7, 27.9, 26.6\\
LSST Y3 & $ugrizy$ & - &   25.6, 26.8, 27.0, 26.3, 25.5, 24.2\\
LSST Y10 & $ugrizy$ & - &  26.3, 27.5, 27.7, 27.0, 26.2, 24.9\\
\hline
\end{tabular}
\end{center}
{
\small
$^\dagger$SSP Deep field sensitivities include those of the Subaru HSC data and of the CLAUDS $u$-band data from CFHT MegaCam \citep{sawicki19}. $^\ast$For SHELA, we give the RA range instead of the pointing center.
{\bf DDF:} LSST Deep Drilling Field ({\tt https://www.lsst.org/scientists/survey-design/ddf}); {\bf SSP:} Hyper SuprimeCam Subaru Strategic Plan ({\tt https://hsc.mtk.nao.ac.jp/ssp/survey}); {\bf EDF:} Euclid Deep Field ({\tt https://www.cosmos.esa.int/web/euclid/euclid-survey}); {\bf HETDEX:} the Hobby-Eberly Telescope Dark Energy eXperiment ({\tt https://hetdex.org}); {\bf SPHEREx} ({\tt https://spherex.caltech.edu/})} 
\label{tab:fields}
\end{table*}

\begin{figure*}
\centering
\includegraphics[scale=0.51]{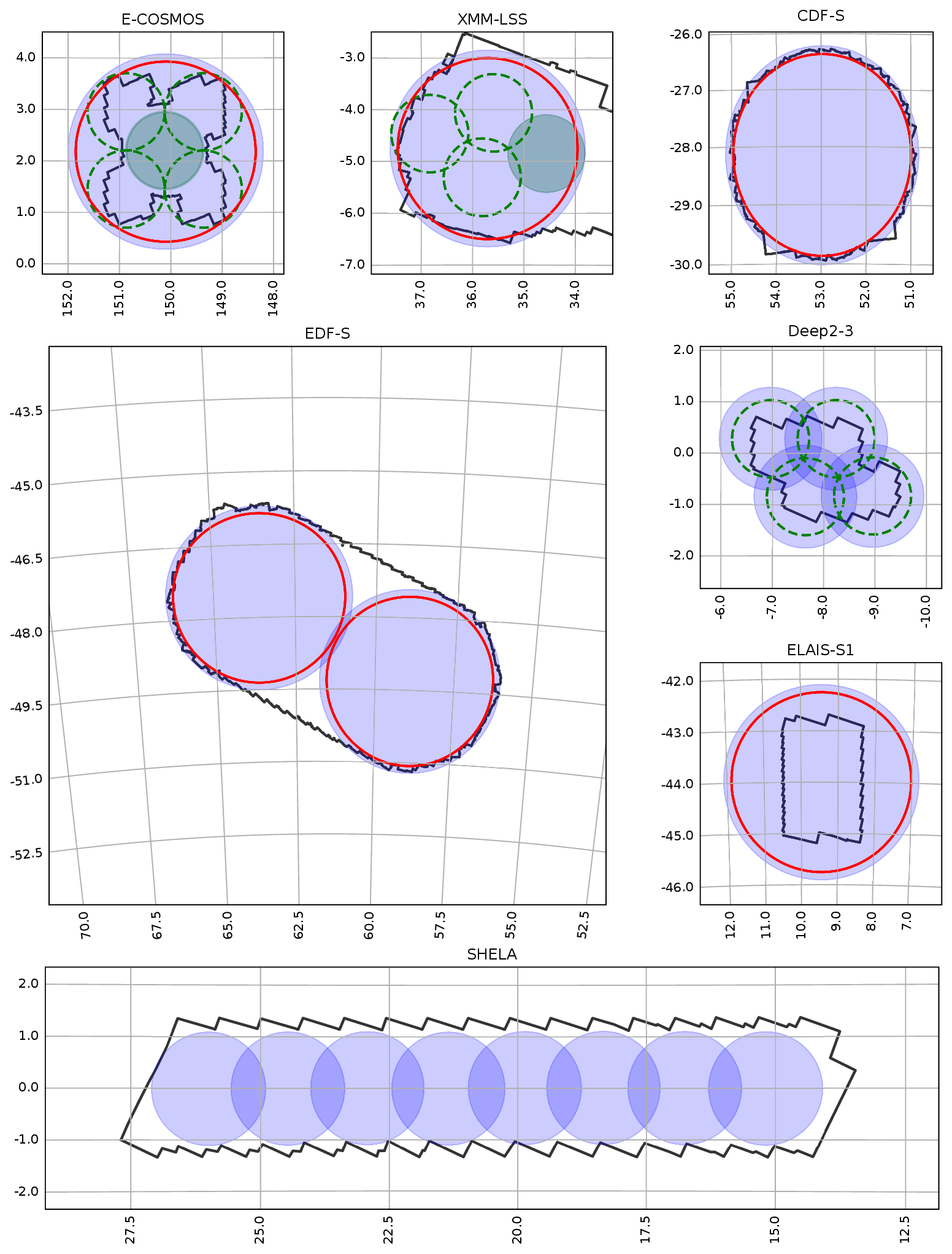}
\caption{The layout of the seven ODIN fields. Blue shades indicate the area covered by ODIN  at  $\geq 70$\% of the target depths. Red circles mark the LSST field of view. Green solid (dashed) circles denote the SSP ultradeep (deep)  regions. All seven fields will receive LSST coverage. 
%The Euclid and HETDEX coverage is indicated by grey solid lines. 
The existing {\it Spitzer} IRAC coverage is outlined in grey. }
\label{fig:fields}
\end{figure*}

The choice of  ODIN fields is based primarily on the maximum depth of existing or future broad-band imaging data, and in several cases,  on the availability of spectroscopy. The seven ODIN fields include three of the Subaru Strategic Program deep fields, all five LSST Deep Drilling Fields (DDFs), two Euclid Deep Fields, and two fields with (nearly) full coverage of spectroscopic coverage from the Hobby-Eberly Telescope Dark Energy eXperiment (HETDEX). Together, these datasets will enable a wide range of scientific investigations on the properties of LAEs, the utility of LAEs as tracers of the large-scale structure, and identification of protoclusters and their signposts  (Section~\ref{sec:odin_science}). 
The complete list of the ODIN survey fields and their characteristics are presented in Table~\ref{tab:fields} and the layout of each field is summarized in Figure~\ref{fig:fields}. 

Within a year after the LSST begins, five of the seven ODIN fields will have the deepest wide-field optical data available (Table~\ref{tab:fields}), surpassing the current depths of the SSP Deep fields. Three of the SSP deep fields are also ODIN fields, namely E-COSMOS, XMM-LSS, and Deep2-3, which are the priority fields for ODIN as they enable scientific investigations before the LSST begins. SHELA is one of the two fields being observed with a near 1:1 fill factor by the integral field spectrographs of HETDEX.
Although SHELA is not part of any deep field imaging surveys, ODIN has been steadily improving on the BB depths in the area that offers continuous IRAC coverage. 

\subsection{Survey parameters and expectations} 

The ODIN survey parameters are optimized for the detection of protoclusters as LAE overdensities. The ability to detect protoclusters as medium-scale enhancements above the average field depends on a combination of two factors. First, a high surface LAE density, $\Sigma_{\rm LAE}$, improves the significance of protocluster detection by lowering the Poisson fluctuations associated with galaxy counts. Second, a smaller line-of-sight thickness ($d_{\rm los}$) minimizes the projection effects that typically suppress, but sometimes enhance, measured overdensities. To this end, the detection efficiency parameter $p_p \equiv \Sigma_{\rm LAE}/d_{\rm los}$ is kept relatively constant at all three ODIN redshifts. However, the DECam Exposure Time Calculator (ETC) suggested that achieving the same $p_p$ level for $N673$ requires prohibitively large integration times ($\approx$ 50\% more than $N419$ and $N501$ time combined). We therefore choose a slightly lower $p_p$ value to set the nominal depth for $N673$ observations.

While $d_{\rm los}$ is tied to the width of the filter transmission, the mean LAE surface density, $\Sigma_{\rm LAE}$, depends on the imaging depth. 
Because the broad-band data typically extend $\gtrsim$1~mag deeper than our narrow-band observations, $\Sigma_{\rm LAE}$ is primarily determined by the seeing and our narrow-band filter exposure times. We estimate the LAE surface density at a fixed NB depth
 using field Ly$\alpha$ LFs \citep{gronwalletal07,ouchi08,ciardulloetal12,sobral18} interpolated to the redshift corresponding to each narrow-band filter's central wavelength.  In this estimate, we assume that the Ly$\alpha$ luminosity dominates the narrow-band flux density.

\begin{table*}
    \caption{Expectations for the ODIN survey}
%    \centering
\begin{center}
    \begin{tabular}{lcccc}
    \hline
     & $N419$ & $N501$ & $N673$ & Total \\
    \hline\hline
     Redshift & $2.45\pm0.03$ & $3.12\pm0.03$ & $4.55\pm0.04$ & -\\
     Exp. Time [hr] & 6.0 & 5.0 & 9.0 & -\\
     $5\sigma$ depth$^\dagger$ [mag] & 25.5 & 25.7 & 25.9 & -\\
     $5\sigma$ Line Flux [cgs$^\ast$] & $3.1\times10^{-17}$ & $1.8\times10^{-17}$ & $1.1\times 10^{-17}$ & -\\
     SB limit [cgs$^\ast$] & $3.5\times 10^{-18}$ & $2.1\times 10^{-18}$ & $1.2\times 10^{-18}$ &-\\
     \hline
     Line-of-sight thickness [cMpc] & 76.0 & 57.0 & 50.6 & - \\
     Cosmic Volume [cMpc$^{3}$] & $7.43\times 10^7$ & $7.04\times 10^7$ & $8.50\times10^7$ & $2.30\times 10^8$ \\
     LAE density [arcmin$^{-2}$] & $0.19$ & $0.17$ & $0.10$ & $0.46$\\
     Protocluster selection efficiency, $p_p$ & $1.1\times 10^{-3}$ & $1.3\times 10^{-3}$ & $0.8\times 10^{-3}$ & - \\
     \# of LAEs & 60,000 & 50,000 & 30,000 & 140,000\\
     \# of protoclusters$^\ddagger$ & 14 (184) & 13 (175) & 16 (211) & 42 (570) \\
     \# of LABs & 268--1874 & 254--1775 & 308--2154 & $\approx$800--5800\\
     \hline
    \end{tabular}
    \end{center}
    $^\dagger$The limiting magnitudes are computed within 2\arcsec\ diameter apertures assuming $1\farcs2$, with seeing (and aperture) scaled with wavelength as predicted in the DECam ETC. We have assumed an average sky brightness corresponding to 0, 3, and 3 days from new moon for $N419$, $N501$, and $N673$, respectively.\\
$^\ast$Line fluxes and surface brightness limits are given in units of erg~s$^{-1}$~cm$^{-2}$ and erg~s$^{-1}$~cm$^{-2}$~arcsec$^{-2}$.\\
$^\ddagger$The number of Coma (Virgo) analogs expected within the survey volume.\\ 
%$^{\ast\ast}$Expected number of Coma (Virgo) progenitors within the survey volume with present-day masses $\gtrsim$$10^{15}M_\odot$ ($(3-10)\times 10^{14}M_\odot$), respectively, scaled from \citet{chiang13}.\\
%$^\ddagger$Expected number of LABs within the survey volume assuming a constant space density $2.5\times 10^{-6}$~cMpc$^{-3}$. The range reflects the current measurement uncertainties and strong field-to-field variations in the LAB number count from existing surveys sampling much smaller volumes \citep{yang10}.
    \label{tab:odin_summary}
\end{table*}

\begin{figure*}[htb!]
\centering
\includegraphics[scale=0.25]{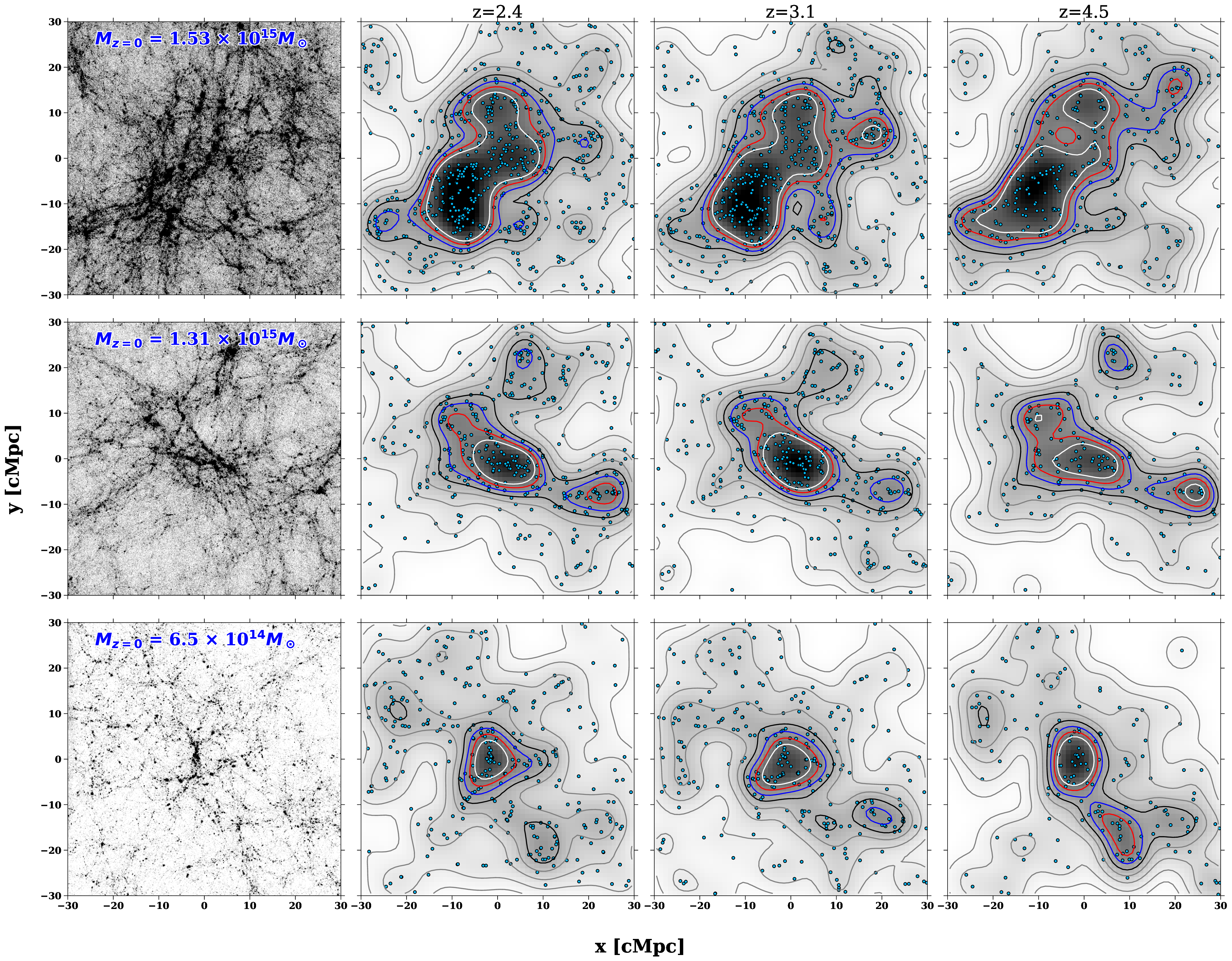}
\caption{
On the leftmost column, we show the dark matter distribution at $z=3$ around three present-day cosmic structures identified in the IllustrisTNG TNG300 simulation. In all cases, the line-of-sight thickness is 60~cMpc, roughly matched to the ODIN survey. The present-day masses are also indicated. Of the halos hosting galaxies with stellar mass $M_{\rm star}\geq 10^9M_\odot$, we randomly assign a subset as `LAEs' to match the expected field LAE surface density at $z=2.4$, 3.1, and 4.5. The mass threshold is chosen to reproduce the observed LAE clustering strengths. The second, third, and fourth columns show the positions of these mock LAEs together with the Gaussian-kernel-smoothed density maps. The white, red, blue, and black lines denote $5\sigma$, $4\sigma$, $3\sigma$, and $2\sigma$ iso-surface-density contours, respectively. These images build clear expectations of how LAEs can trace massive cosmic structures at high redshift.
}
\label{fig:tng_protoclusters}
\end{figure*}

In Table~\ref{tab:odin_summary}, we list the nominal depths of our narrow-band data and the corresponding Ly$\alpha$ flux and the surface brightness limits. The expected LAE surface density and number of LAEs are also summarized. We also list the comoving line-of-sight thickness, the comoving cosmic volume, and the protocluster detection efficiency ($p_p$). In order to calculate the cosmic volume, we assume the full survey area of 91~deg$^2$. The expected number of protoclusters for each redshift slice is estimated based on this volume and the comoving number densities of clusters. As for the number densities of Coma- and Virgo-sized clusters (present-day masses of $\geq 10^{15}h^{-1}M_\odot$ and $(3-10)\times 10^{14}h^{-1}M_\odot$, respectively), we assume $1.84\times 10^{-7}$ and $2.48\times 10^{-6}$~cMpc$^{-3}$, respectively. These numbers are taken from the Millennium simulations-based \citep{springel05} values computed by \citet{chiang13}. The adopted cosmology for the simulation\footnote{The cosmological parameters assumed for the Millennium Run is $\Omega=0.25$, $\Omega_\Lambda=0.75$, $h=0.73$.}  is only slightly different from ours and we do not make any additional corrections. The uncertainty due to Poisson shot noise and cosmic variances are expected to be larger than the changes from different cosmologies, particularly for the Coma progenitors. 

As for the expected number of LABs, we assume a constant density across all redshifts. The number density estimate for bright and extended LABs  \citep[$L_{{\rm Ly}\alpha}\gtrsim 1.5\times 10^{43}$~erg~s$^{-1}$, $A_{\rm iso}\geq 16$~arcsec$^2$:][]{yang10} is $1.0^{+1.8}_{-0.6}\times 10^{-5}$~cMpc$^{-3}$ adopting a cosmology with $\Omega=0.3$, $\Omega_\Lambda=0.7$, and $h=0.7$. The uncertainties reflect strong field-to-field variations as well as the uncertainties originating from shot noise for a small-area (1.2~deg$^2$) survey. The values listed in Table~\ref{tab:odin_summary} are corrected for the difference in cosmology which is at a $\approx$10\% level. Recently, from the ODIN COSMOS/$N501$ data, \citet{ramakrishnan23} reported 129 LABs over a $\approx$9~deg$^2$ region, forecasting that the total number of LABs may be closer to the upper limit in the table. More stringent limits will be placed by combining the LAB samples in multiple fields (B.H. Moon et al., in preparation).

In Figure~\ref{fig:tng_protoclusters}, we illustrate how massive structures may appear at the nominal $p_p$ range adopted by ODIN. Using the IllustrisTNG simulation suite \citep{weinberger17}, we first choose three massive structures. In terms of $M_{z=0}$, the total mass at $z=0$, they are ranked \#1, \#2, and \#7, 
%first, second, and 7th with the total masses of 15.3, 13.7 and 6.5 $\times$ $10^{14}M_\odot$, 
respectively. The dark matter distribution of these structures at $z=3$ is shown in the left panels. To match the survey parameters, we create a (60~cMpc)$^3$ volume centered on the progenitor of each cluster at $z=3$. Adopting a minimum stellar mass of $M_{\rm star}=10^9~M_\odot$ to match the LAE clustering measurements from the literature \citep{gawiser07,kovacetal07,guaita10,kusakabe18}, we randomly select a subset of halos above the threshold that matches the ODIN surface densities (Table~\ref{tab:odin_summary}). The surface density map is then created by smoothing the `LAE' distribution with a fixed kernel. We choose a two-dimensional Gaussian kernel with a full-width-at-half-maximum of $5\farcm2$ (10~cMpc at $z=3$), comparable to the expected size of a protocluster. We note that, as a simple demonstration, we use the $z=3$ matter distribution in these structures to simulate the LAE maps at all three ODIN redshifts by simply matching the respective LAE surface densities. 

Figure~\ref{fig:tng_protoclusters} demonstrates that ODIN is expected to detect Virgo- or Coma progenitors with ease provided that LAEs represent an unbiased tracer of low-mass halos. While we chose to use a random subset of halos as mock LAEs, more realistic modelings of LAEs following the recipes of \citet{weinberger19} or \citet{mason18} do not meaningfully alter the result \citep[][M. Andrews et al., in prep, S. Im et al., in prep]{lee23}.
However, reality may be more complicated. Radiative transfer effects could cause LAEs to avoid the regions of highest density \citep[e.g., see][]{momose21,huang22}. Additionally, as galaxies grow more massive, star formation activity is expected to ramp up and produce more interstellar dust, lowering their likelihood of being classified as LAEs \citep[e.g.,][]{weiss21}. These considerations suggest that the densest and most evolved regions may not be represented as the highest LAE overdensities. 
%LAE-traced densities may not fully represent the true density field in its densest and most evolved regions. 
More observational evidence is needed to firmly establish the level of this potential bias. 

\begin{figure}
\centering
\includegraphics[scale=0.6]{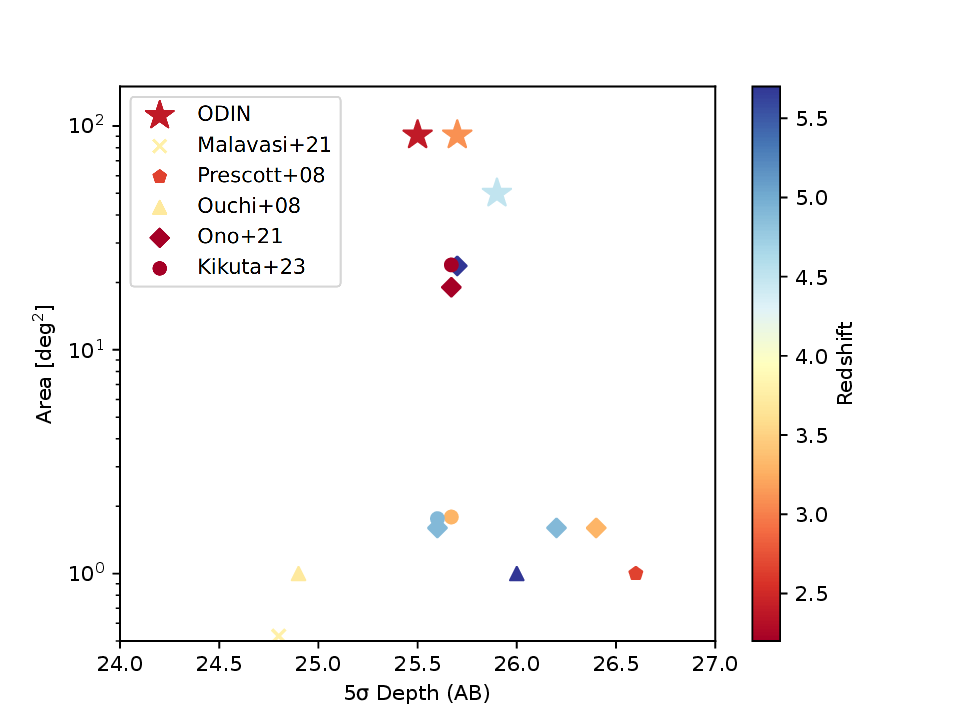}
\caption{Comparison of the imaging depth and area coverage of recent surveys with ODIN (stars). All data are color-coded by redshift. Diamonds and circles represent the combined SILVERRUSH and CHORUS surveys \citep{Ono:2021,kikuta23}. Earlier surveys with Subaru/SuprimeCam \citep[][triangles and pentagon, respectively]{prescott08,ouchi08}, Mayall/Mosaic \citep[][cross]{malavasi21} are also listed.
}
\label{fig:compare_survey}
\end{figure}

In Figure~\ref{fig:compare_survey}, we show the areal coverage vs. depths of existing narrow-band $z$=2--5 Ly$\alpha$ surveys along with the ODIN survey. These studies include the combined results from the SILVERRUSH and CHORUS conducted with Subaru/HSC  \citep{Ono:2021,kikuta23} as well as several earlier measurements with Subaru/SuprimeCam \citep{ouchi08,prescott08} and Mayall/Mosaic3  \citep{malavasi21}. 

\section{Observations and Data Reduction} \label{sec:obs}

\begin{figure*}
\centering
\includegraphics[scale=0.9]{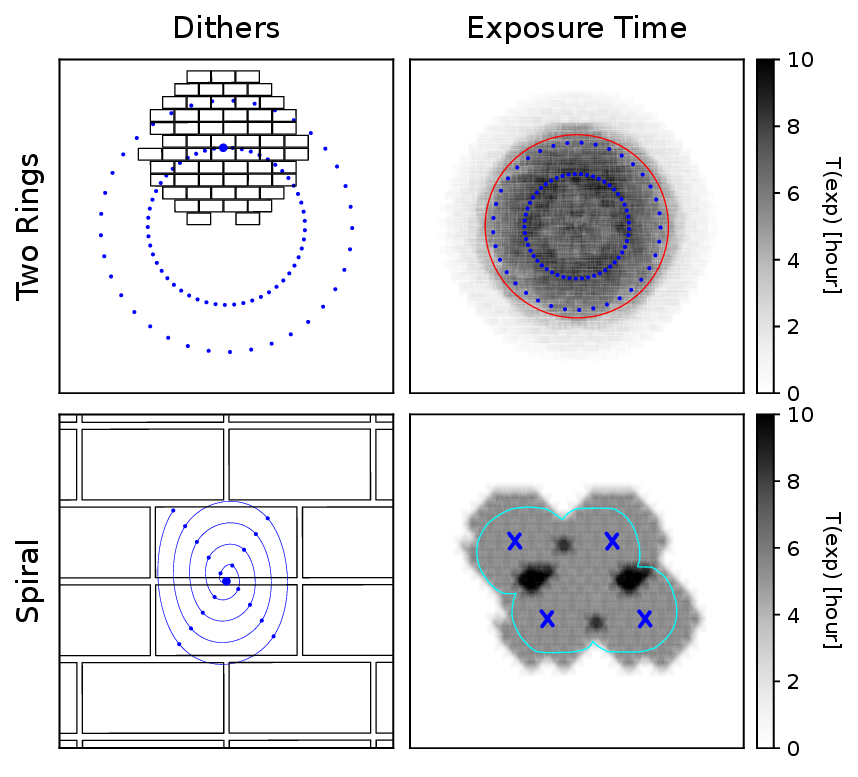}
\caption{Two dither patterns used for the ODIN survey. {\it Top left:} Blue points mark the center position of each DECam pointing in the two-ring dither pattern. The radius of the inner and the outer ring is 1.0$^\circ$ and 1.6$^\circ$, respectively. The individual CCDs of DECam are shown in black. 
{\it Top right:} The resultant exposure map for $N419$ together with the pointing centers (blue points) and the LSST field-of-view (red circle). 
{\it Bottom left:} The spiral dither positions are overlaid on the center of the DECam focal plane, with individual CCDs shown in black. Small dithers along the spiral create an X-shaped set of pointings that enable uniform coverage by minimizing the overlap between chip gaps. 
{\it Bottom right:} The exposure map for $N419$ of the Deep2-3 field employs the spiral dither pattern of four overlapping fields. The pointing centers are shown in blue. The broad-band coverage from SSP is shown in cyan. 
}
\label{fig:dither}
\end{figure*}

\subsection{Dither patterns}\label{subsec:dither}
DECam images a circular region of 2.1$^\circ$ in diameter. In comparison, the LSST field-of-view (FOV) is 3.5$^\circ$ in diameter. To achieve a reasonable level of uniformity across the five LSST DDFs using DECam, we designed a two-ring dither scheme as illustrated in the top left panel of Figure~\ref{fig:dither}. DECam pointings are positioned along two concentric rings with radii 1.0$^\circ$ and 1.6$^\circ$ from the field center.\footnote{Although this is being described as a dither pattern consisting of two rings of large dithers without ever using the central pointing, it could equally well be described as a two-ring pointing pattern with no dithers performed.}  The number of visits in each ring is determined based on the 1200~s exposure time per frame and the effective exposure time per pixel needed to reach the target depths as discussed in Section~\ref{subsec:reduction}. The radii and total exposure time ratio of the rings are calibrated to minimize the variation of the depths in the radial direction by controlling the degree of overlap between the inner- and outer-ring exposures. 
%between the exposures taken as part of the inner and those for the outer rings. 

In the top right panel of Figure~\ref{fig:dither}, we show the nominal exposure time map for the $N419$ observations in COSMOS. The final map is similar to this figure except for the presence of satellite trails. Each sky location within the LSST Deep Drilling Field gets observed by at least 10 pointings, naturally filling in chip gaps to yield locally uniform depth.  The angular spacing between adjacent pointings in a given ring is smaller for the $N501$ and $N673$ observations because the exposure time per pixel is higher, but adheres to the same principle. The exposure maps of all DDF fields in all filters are expected to be similar to that shown in the figure. 

In a given ring, we randomly shuffle the order in which pointings are observed to ensure uniformity in seeing across the field. We typically complete the inner ring pointings before starting the outer ring to enable early science. For the EDF-S, which is composed of two slightly overlapping LSST pointings, we randomly select pointings from rings in both EDFS-a and EDFS-b, as the CTIO 4m slew time from one to the other is never much larger than the DECam readout time.  

For the fields that are not DDFs, namely, SHELA and Deep2-3, we take a different approach. For SHELA, we match the NB coverage as closely as possible to the existing BB data  \citep{wold19}. The coverage is comprised of eight DECam pointings that follow a spiral pattern with small dithers to fill gaps, as shown in Figure~\ref{fig:fields}. On the other hand, Deep2-3 is defined via four overlapping HSC pointings. In both SHELA and Deep2-3, we employ a spiral dither pattern.  As illustrated in the bottom left panel of Figure~\ref{fig:dither}, dither positions move away from the pointing center along a spiral by a small offset each time such that chip gaps along the RA and Dec axes never repeat, yielding an X-shaped pattern designed to fill the CCD gaps and to cover the BB regions with uniform coverage.
In all narrow-band filters, we use the identical pattern repeating it as many times as needed to achieve the target depth.

\subsection{ODIN data acquisition and reduction}\label{subsec:reduction} 

%- a paragraph about how individual exptime was determined

%- a few paras about data reduction and calibration

%Individual DECam frames are processed and coadded with the DECam Community Pipeline \citep{Valdes2014,DCP} into a single image.

The ODIN survey began its observations in February 2021. At the time of this article, the survey is $\approx$50\% complete. A comprehensive summary of the LAE selection will be given elsewhere (N. Firestone et al., in preparation), and here we focus on the general practice adopted for our observations and data processing. 

ODIN observations of all filters use an exposure time of 1200~sec, which is motivated by two considerations: the low sky background expected in narrow-band filters requires that our exposure times be long enough to ensure that the 7~e$^-$ read noise of the DECam CCDs is not a dominant factor. 
%Readout noise of DECam is 7~e$^-$ per pixel. 
According to the DECam exposure time calculator, in a 1200~sec exposure in $N419$, $N501$, and $N673$, the fraction of readout noise is $
\approx$30, 13, and 4\% of the sky background noise measured in a 2\arcsec\ diameter circular aperture at new moon.
%, respectively, in a 2\arcsec\ diameter circular aperture.  
A countervailing need is to avoid filling too much of the detector with cosmic rays, which, given the thick CCDs on DECam, can be highly extended and morphologically complex. This drives us to set a maximum exposure time of 1200s.  

During our observations, we utilize the {\it copilot} software \citep{copilot}, a  tool developed for DECam to monitor observing conditions and image weight. Using the Pan-STARRS-1 photometric catalog \citep{2012ApJ...756..158S}, copilot measures seeing, transparency ($T$), and sky brightness ($B$). Based on these data, we compute relative image weights, $w$, as: 
\begin{equation}\label{eq:eq1}
 w \propto \frac{T^2 ~{\rm {\tt exptime}}^2}{B {\rm~{\tt seeing}}^2} 
\end{equation}
The weight is normalized by values expected in a nominal dark sky condition. For ($N419$, $N501$,$N673$) a weight of unity would be achieved in a 1200~sec exposure with $T=1$, seeing of (1.2,1.1,1.0)\arcsec\ and a background level of (50,100,200)~ADU per pixel. The primary goal is to ensure that each designated pointing receives a combined weight of $\approx$1 by the completion of our observations. In practice, this means that the same pointing can be revisited to add more depth until the combined weight is similar to the original specification. The weighting scheme given in Equation~1 can accommodate shorter-exposure frames as long as readout noise remains sub-dominant.

We process and stack the ODIN data using the DECam Community Pipeline \citep[CP:][]{Valdes2014,DCP} as follows. %As a detailed description of data reduction is given in \citet{ramakrishnan23}, we provide a brief summary here. 
%The $N501$ data consists of 72 individual DECam exposures (each with exposure time of 1200~s)  taken in February 2021; the total  observing time is 24~hrs for the field with a per pixel exposure time range of 20~min to 7.3~hrs and average of 2.9~hrs or 3.1~hrs (minimum of 1 or 2 overlapping exposures respectively). Individual DECam frames are processed and coadded with the DECam Community Pipeline \citep{Valdes2014,DCP} into a single image.
Each DECam exposure consisting of data from the 60 calibrateable CCDs is flat-fielded separately by dome flats, star flats, and dark sky illumination flats.  Dark sky illumination flats are created by coadding unregistered stacks of exposures. Dome flats are produced by stacking sequences of 11 exposures taken each night; star flats are made  from widely dithered exposures of a star field 
%with many bright stars 
using the algorithm of  \citet{2017PASP..129k4502B}. 

For each CCD, the background is measured by the modes of pixel values in small blocks with sources masked. 
%where the pixels belonging to astronomical sources are rejected using iterative clipping. 
%We fit a low-order polynomial  to the modes and uniformize the background by subtracting it from each CCD.
Blocks with an insufficient number of background pixels are replaced by interpolation from nearby blocks. A spline surface is fit to the modes. In fields with prominent reflections from very bright stars, the modes are used directly to best handle the fine structure of the ghosts. The background is subtracted while maintaining the mean. 
%The background is then made uniform by matching the means in each CCD and subtracting a low-order fit to the modes. 
The block size chosen to measure the background is sufficiently large to ensure that it does not introduce any photometric bias for extragalactic sources; however, 
since the size is smaller than the region affected by bright stars, the procedure can lead 
%it is still smaller than the regions affected by bright stars, which leads 
to over-subtraction of the faint halos. 

The CP provides a data quality flag for every pixel for every exposure. The flags, in order of precedence, are known bad CCD pixels, saturated or bleed trail pixels, and cosmic ray and streak identifications. The last analysis is done on individual exposures but multiple ODIN exposures allow further identification via transient signals. 
In addition, the ODIN team produces satellite trail masks for affected exposures which are combined with the CP data quality masks. 
%Satellite trail masks are created at this stage and combined with quality masks.
%If an externally supplied data quality mask is available for an exposure, such as the satellite trail masks produced by the ODIN team, it is added to the CP data quality mask for that exposure. 
%In the creation of the ODIN deep field stacks, the pixels in the data quality mask are excluded from consideration. 
The pixels flagged in the master quality mask are excluded from consideration when coadding all the exposures to create the deep image stacks. 

%While this step is critical to producing a uniform dithered stack, it leads to over-subtraction of the faint halos around bright stars.  However, we remove any science source close to bright stars in the analysis by applying star masks (Section~\ref{subsec:sources}). Thus, the effect of uneven background levels near bright stars on the small, distant extra-galactic objects is negligible. 

An astrometric solution is derived for each CCD by matching stars to Gaia-EDR3 \citep{2021A&A...649A...1G}. While the higher order distortions are predetermined and fixed,  the low order terms are updated using the astrometric solver SCAMP \citep{2006ASPC..351..112B} with continuity constraints between CCDs. The root-mean-square of a plate solution is typically a few hundredths of an arcsecond. We reproject the exposures to a standard tangent plane sampling at the pixel scale of $0\farcs27$/pix using sinc interpolation. A fixed tangent point is used for all the exposures in a given field with the exception of the two largest ODIN fields which are more than 10$^\circ$ across. For EDF-S and SHELA, two and four separate tangent points are used. 

Individual exposures are matched to the Pan-STARRS-1 photometric catalog \citep{ 2012ApJ...756..158S} to determine a zeropoint ({\tt magzero}) and a photometric depth in magnitude ({\tt depth}). The latter is computed as:
%photometric depth in magnitude is computed as:
\begin{equation}
{\tt depth} = {\tt magzero} - 2.5\log \left ( 5\sigma  \sqrt{\pi r^2_{\rm op}} \right )
\end{equation}
where $\sigma$ is the pixelwise sky noise and $r_{\rm op}$ represents the size of an optimal aperture (for a pure Gaussian seeing) whose diameter is 1.35 times the measured seeing FWHM.
%\begin{equation}
%r_{\rm op} = \frac{1.35~ {\tt seeing}}{2}
%\end{equation}
%    
The {\tt magzero} parameter combines sky transparency and exposure time for a given frame while {\tt depth} represents the $5\sigma$ uncertainty.
%in an optimal aperture for a Gaussian point spread function with the measured seeing and standard deviation of the sky. 
The {\tt depth} parameter is then converted to linear relative counts and squared to form a normalized relative weight for each frame. 
%along with seeing and sky brightness estimates, factor into the weighting of the coaddition. The image weight is computed in the same way as Equation~\ref{eq:eq1}
%where transparency  $T$ is related to the difference in zeropoint between a given exposure and the reference frame taken in photometric conditions as $T\equiv 10^{0.4*\Delta ZP}$.  
%The dithered exposures are stacked by averaging registered pixels with statistical rejection (constrained sigma clipping) of outliers to minimize cosmic rays accumulated from the long exposures.
All exposures are coadded as a weighted mean with statistical rejection (constrained sigma clipping) of outliers to remove the large number of cosmic rays accumulated in our long exposures. 

\begin{figure*}
\centering
\includegraphics[scale=0.65]{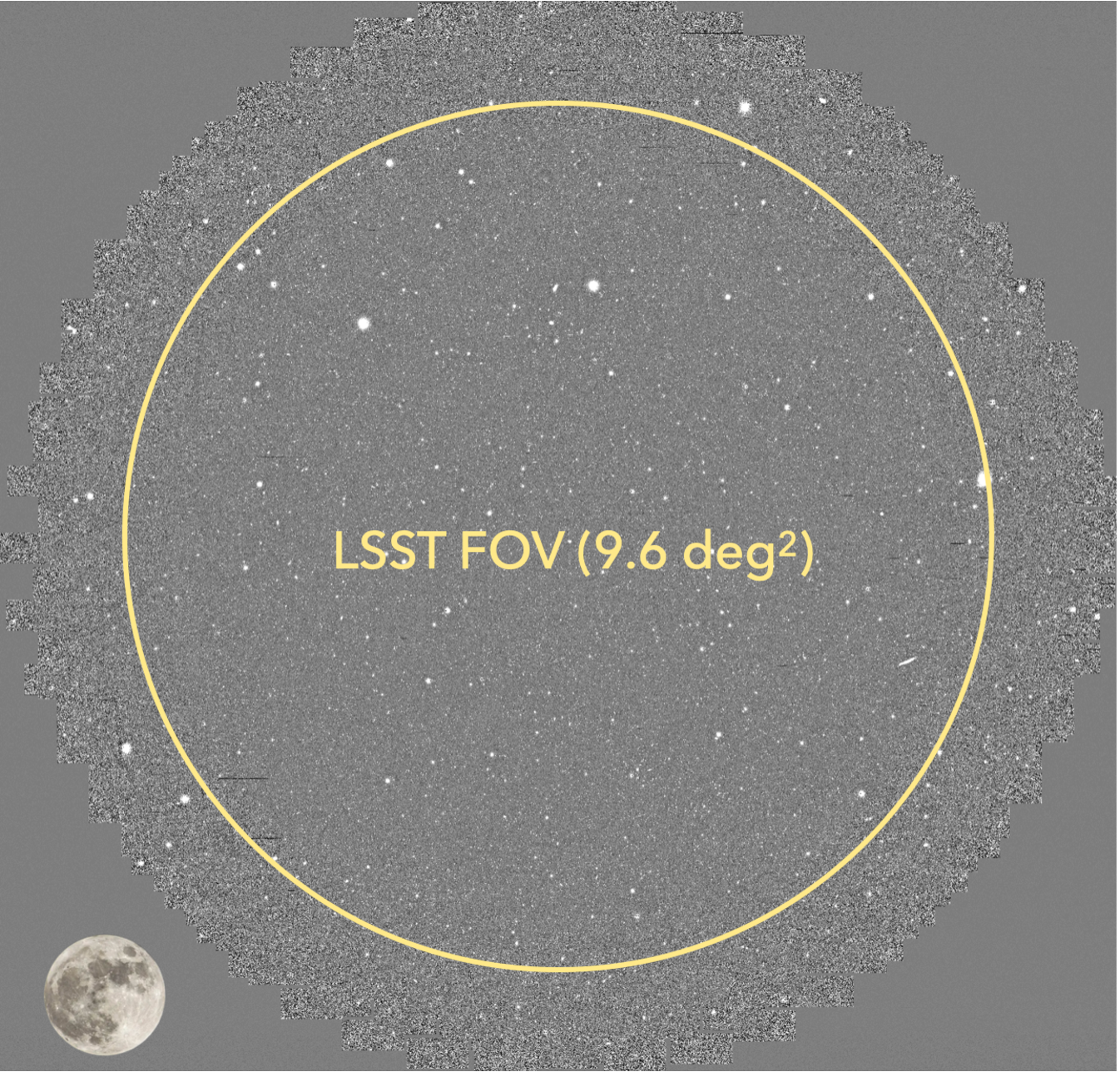}
\caption{
The final mosaic of CDF-S in $N501$ is shown. The yellow circle marks the position and area coverage of the planned LSST observations of CDF-S in $ugrizy$ at the DDF depths indicated in Table~\ref{tab:fields}. The to-scale size of the Moon is shown at the bottom left corner.
}
\label{fig:ecdfs}
\end{figure*}

In Figure~\ref{fig:ecdfs}, we show the full-depth mosaic of CDF-S in $N501$ overlaid with the LSST field-of-view (yellow circle) and the size of the Moon. Similar to the SSP data release, the ODIN stack is split into multiple `tracts', each $1.7^\circ \times 1.7^\circ$ in size with an overlap of 1\arcmin\ for the ease of data handling. The only exception to this convention is SHELA, for which we create four separate stacks each of which contains two adjacent DECam pointings.

%The exposures are matched to the Pan-STARRS-1 photometric catalog \citep{ 2012ApJ...756..158S} for a flux zero point to provide the scaling and, along with seeing and sky brightness estimates, weighting of the coadd. The dithered exposures are stacked by averaging registered pixels with statistical rejection (constrained sigma clipping) of outliers to minimize cosmic rays accumulated from the long exposures.
%The SSP tracts are 
%The BB images from the HSC-SSP and CLAUDS surveys are available in the form of contiguous `tracts' of approximately $1.7^\circ \times 1.7^\circ$, with an overlap of $\sim$ 1\arcmin. These tracts are 1\arcmin. 
%reprojected using the DECam Community Pipeline to have the same tangent points and pixel scales (0.27\arcsec/pixel) as the ODIN data. %In order to facilitate multiband photometry, we divided our NB image to match the BB tracts. 

\section{External Data}\label{sec:external_data}

\subsection{Broad-band optical data} 
The detection of high-redshift LAEs requires that the ODIN narrow-band data be compared existing broad-band frames.
%We utilize the existing broad-band imaging data. 
For the Deep2-3, E-COSMOS, and XMM-LSS fields, these data come from the HSC SSP program \citep{ssp_surveydesign}. Specifically, we use the publicly available frames from the third data release \citep{ssp_dr3}. Similar to that described in Section~\ref{subsec:reduction}, we reproject the data using the common tangent point at the pixel scale of $0\farcs27$/pix. 

In the SHELA field, we use the archival DECam data taken as part of several NOIRLab programs. As described in \citet{wold19}, the existing observations are composed of eight overlapping DECam pointings. We reprocess the data using the DECam CP and find that the depths vary significantly across the field. Since Fall 2021, ODIN has been conducting additional DECam imaging with the primary goal of homogenizing the sensitivity of broad-band data with the top priority on the $gr$ bands to enable a uniform selection of LAEs.

In the remaining three fields (EDF-Sab, ELAIS-S1, and CDF-S), the availability of the BB data across the $\sim10$ deg$^2$ field is currently limited. A subsection of CDF-S was observed repeatedly as one of the Dark Energy Survey deep fields \citep{hartley22} designed to detect supernovae. While the approximate depths of these data are listed in Table~\ref{tab:fields}, the sensitivity will be easily surpassed in 2026 by Rubin with uniform coverage within the ODIN fields. These considerations have been taken into account in setting the observational priorities. 

\subsection{Spitzer IRAC data}\label{subsec:spitzer} 
All seven ODIN fields have existing {\it Spitzer} IRAC images which cover a substantial fraction of the survey area at depths that are useful for selecting massive galaxies in the ODIN redshift range. These datasets will enable statistical association of massive galaxies with cosmic structures and/or Ly$\alpha$ blobs and elucidate the early formation histories of massive cluster ellipticals. The IRAC coverage of the ODIN fields is indicated by grey lines in Figure~\ref{fig:fields}. Here, we briefly summarize the publicly available data. 

The COSMIC DAWN survey \citep{spitzer_euclid} obtained IRAC imaging of three Euclid Deep Fields. In all cases, we quote the $5\sigma$ detection limit, which is $\approx$0.5~$\mu$Jy for EDF-N and CDF-S (i.e., the Euclid Deep Field Fornax) and $\approx$1.0~$\mu$Jy for EDF-S.  Additionally, Cosmic Dawn re-reduced all existing IRAC datasets in other Euclid calibration fields including XMM-LSS and COSMOS (both at $\approx$1.0~$\mu$Jy) in a consistent manner. Smaller subsections of COSMOS, XMM-LSS, and CDF-S have also been imaged to varying depths by other programs. For a summary of the depths and areal coverage, we refer interested readers to  \citet{spitzer_euclid}. 

ELAIS-S1 and Deep2-3 are covered by the Spitzer Extragalactic Representative Volume Survey \citep[SERVS;][]{servs},
%A larger portion of Elais-S1 is covered by the Spitzer Wide-area Infrared Extragalactic survey \citep[SWIRE:][]{swire}; however, the imaging depth of SERVS is more suitable for finding massive galaxies at $z>2$. 
and the Spitzer coverage of HSC-Deep with IRAC for Z \citep[SHIRAZ;][]{shiraz} survey, respectively, at a comparable depth of $\approx$1~$\mu$Jy. A larger section of ELAIS-S1 is imaged at a shallower depth ($\approx$3~$\mu$Jy) by the Spitzer Wide-area InfraRed Extragalactic survey \citep[SWIRE;][]{swire}. Finally, the imaging of the SHELA was performed by the Spitzer-HETDEX Exploratory Large Area survey \citep[SHELA:][]{Papovich2016:SHELA} at the shallowest depth (5.5~$\mu$Jy), which corresponds to a star-forming galaxy of moderate dust reddening, $E(B-V)=0.25$ and stellar mass of $\approx 10^{10.2}M_\odot $ at $z=2.5$. 

\subsection{Spectroscopy} 

Multiple ongoing spectroscopic programs are targeting ODIN-selected galaxies with DESI, Gemini GMOS, Keck DEIMOS, and the Prime Focus Imaging Spectrograph on the South African Large Telescope. These data have been highly successful in validating LAE selection, measuring the spatial structure of several protoclusters, and spectroscopically confirming more than 40 Ly$\alpha$ blobs. Presently, the number of ODIN LAEs with spectroscopic follow-up is $\approx$8,000, 3,000, and 800 at $z=2.4$, 3.1, and 4.5, respectively. The results from these programs will be presented in forthcoming papers. 

%Through a collaborative agreement, the Dark Energy Spectroscopic Instrument (DESI) collaboration has obtained spectroscopy of ODIN targets. Thus far, $\approx$12,000 ODIN-selected LAEs have been observed in XMM-LSS and COSMOS. These targets typically have line fluxes $\gtrsim 5\times 10^{-17}$~erg~s$^{-1}$~cm$^{-2}$. Several spectroscopic programs have been conducted targeting fainter LAEs using Keck DEIMOS and Gemini GMOS. These programs will help validate the LAE selections and pursue key science investigations of the ODIN survey. More details will be presented in several forthcoming papers. 

%\subsection{The Hyper Suprime-Cam SSP Data}
%- describe the SSP/CLAUDS data and depths here

%\subsection{Spitzer IRAC data}
%- describe the Spitzer IRAC surveys of the fields (their layout can be added to one of  the figures)

%\subsection{Spectroscopy} 
%- the prospects of joint analysis with HETDEX/DESI 

%\section{Results} 
%
%- not sure what to put in here as  Vandana's paper will show the LAE selection, angular distribution, and protocluster selection, etc. 
%
%- Refer to Vandana's  paper for preliminary LAE selection and show redshift distribution, a rough estimate of possible contamination rate, etc. 
%
%- can we show stacked DESI spectra provided that MoU is signed by this time, or is it part of a plan for the DESI-ODIN spectroscopy paper?
%

\section{Ancillary Science}\label{sec:other_applications}

%KSL: I think it is best to not mention any results from primary ODIN science here. What else can we highlight here? 
%EG I agree so altered the section heading.  The spiral galaxy in COSMOS gives a nice flavor for non-primary science with pretty pictures! Sean's image would be good too, along with any o3 nebulae in ECDF-S.   

%. As for the latter, we use the ODIN NB data as `blue' such that any sources with strong emission lines falling into the filter set would appear bluer than the others. 
%In the top panels of Figure~X, we show NGC~XXX. 

\begin{figure*}
\centering
\includegraphics[scale=0.30]{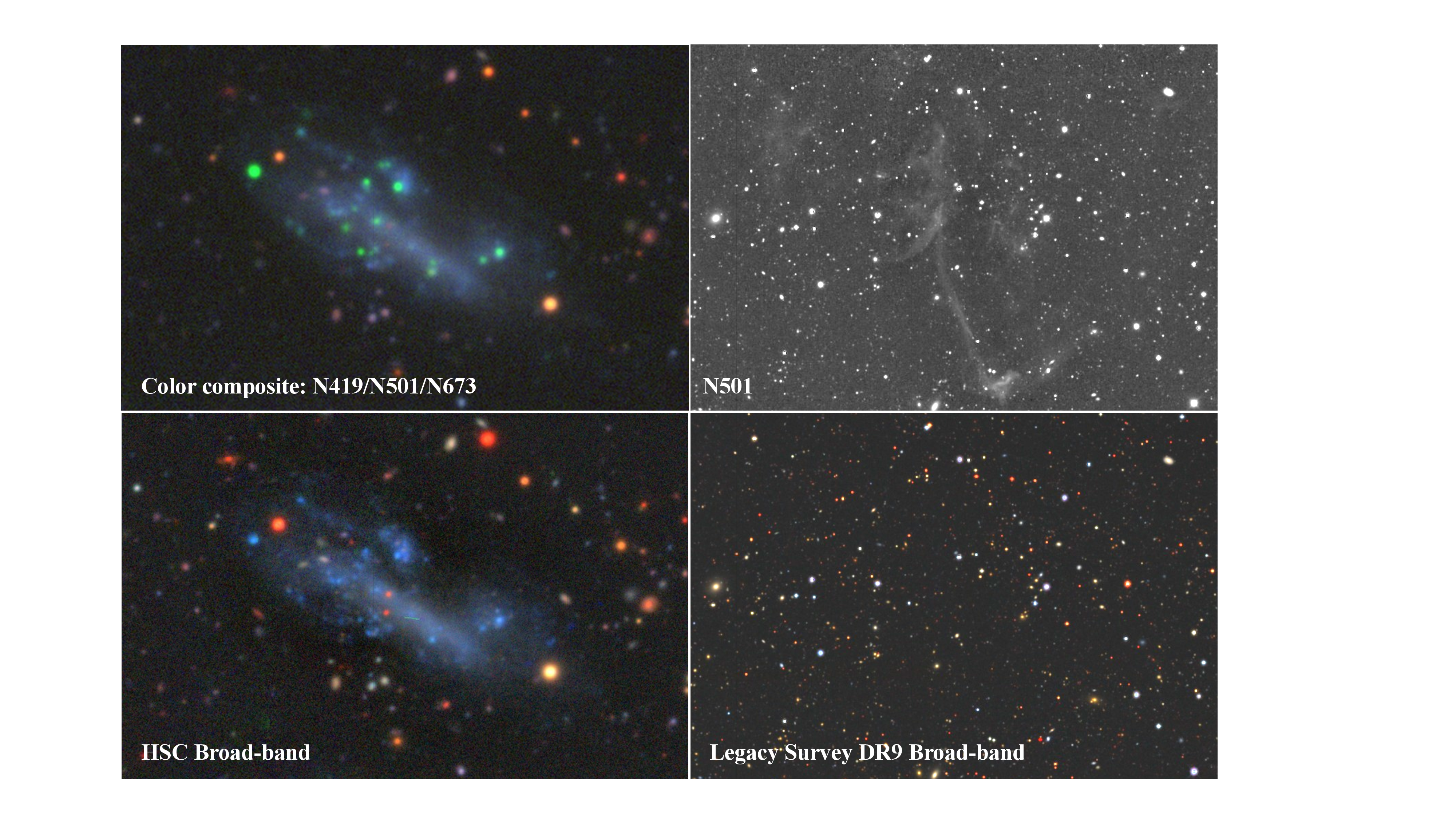}
\caption{{\it Left panels}: A color-composite image of a nearby star-forming galaxy is displayed using the ODIN NB filters (top) and the HSC broad-band filters (bottom). Star-forming knots are clearly delineated in the former in green as $N501$ samples O~{\sc iii} nebular emission. {\it Right panels:} Extended large-scale features, visible in the $N501$ image (top), likely trace shocked, [O~{\sc iii}]-emitting gas associated with a planetary nebula, NGC~1360 located outside the ODIN field.   }
\label{fig:highlights}
\end{figure*}

The deep NB data taken as part of ODIN can be used for a number of programs.  These data can be visualized from the ODIN Legacy Viewer\footnote{\tt https://odin.legacysurvey.org/} where users can browse each stacked ODIN narrow-band image as well as a color composite frame that combines the ODIN NB data together.

The left panels of Figure~\ref{fig:highlights} show a nearby star-forming galaxy located at $(\alpha,\delta) = (149.620^\circ,+1.694^\circ)$. The colors are computed from ODIN data (top) and HSC broad-band images (bottom). Star-forming knots with strong [O~{\sc iii}] emission are clearly delineated as green features in the top panel. 
The right panels show a small section in CDF-S centered on $(\alpha,\delta)=(53.164^\circ, -26.219^\circ)$ observed in $N501$ (top right) with Legacy Survey DR9 broad-band data (bottom right). The complex and very extended [O~{\sc iii}] emission feature likely originates from shocked gas associated with NGC~1360, a planetary nebula located at $(\alpha,\delta)=(53.311^\circ, -25.870^\circ)$, $\approx$20\arcmin\ northeast and outside the ODIN field. In the legacy viewer\footnote{{\tt https//legacysurvey.org}}, the gas is captured in the broad-band data 
%oriented towards the feature 
but only very close to NGC~1360. The sensitivity of the $N501$ and $N673$ data afforded by ODIN provides a unique opportunity to explore faint gaseous components of our Galaxy and star-forming regions of the nearby universe \cite[e.g.][]{m31_o3}.

\section{Summary} 

%In preparing for the upcoming era of deep wide-field surveys such as LSST, Euclid, and Roman, 
The ODIN collaboration is currently conducting a narrow-band survey with the largest area, 
%the widest-field narrow-band imaging campaign, 
covering $\approx90$~deg$^2$ with three custom filters down to depths of 25.5--25.9~AB. By sampling redshifted Ly$\alpha$ emission at $z=2.4$, 3.1, and 4.5 in narrow cosmic slices ($\approx 60$~cMpc in thickness) within a 0.25~cGpc$^3$ volume, ODIN will enable robust measurements of the epochs' large-scale structure by mapping out knots, groups, filaments, and cosmic voids within the underlying matter distribution using low-mass, Ly$\alpha$-emitting galaxies. 

ODIN will discover hundreds of distant protoclusters and thousands of the largest and most luminous Ly$\alpha$ nebulae, allowing us to study the physical association of these two rare types of cosmic objects and their relationships to the surrounding LSS. Through measurements of the LAE luminosity functions, spectral energy distribution, and clustering, we will determine the physical properties of these objects and their association with dark matter halos and obtain quantitative measurements of how LSS influences galaxy formation. The models of expansion history will be explored through measurements across cosmic time. 

With Rubin, Euclid, and Roman on the horizon, the  ODIN fields will become the deepest fields rich with spectroscopic and imaging data. When combined with these data, the ODIN data will have a lasting legacy in advancing studies of not only galaxy evolution at high redshift but also the Galaxy and the nearby universe. 

%the wavelengths of rest-frame [O~{\sc iii}] and [S~{\sc ii}] emission, 

\begin{acknowledgements}
    
Based on observations at Cerro Tololo Inter-American Observatory, NSF's NOIRLab (Prop. ID 2020B-0201), which is managed by the Association of Universities for Research in Astronomy under a cooperative agreement with the National Science Foundation.
%%%%
KSL and VR acknowledge financial support from the National Science Foundation under Grant Nos. AST-2206705 and from the Ross-Lynn Purdue Research Foundation Grant. 
%%%
This material is based upon work supported by the National Science Foundation Graduate Research Fellowship Program under Grant No. 2233066 to NF.  NF and EG 
acknowledge financial support from the National Science Foundation under Grant Nos. AST-2206222 and from the NASA Astrophysics Data Analysis Program grant 80NSSC22K0487.
HS acknowledges the support of the National Research Foundation of Korea grant, No. 2022R1A4A3031306,  funded by the Korean government (MSIT).
LG, JN, and AS acknowledge recognition from Fondecyt Regular no. 1230591, and JN acknowledges support from Universidad Andr\'es Bello grant no. DI-07-22/REG.

\end{acknowledgements}

\bibliography{myrefs}{}
\bibliographystyle{aasjournal}



\end{document}